\documentclass[a4paper,12pt]{article} 
\usepackage{amsmath,amsthm,amssymb}
\newcommand{\bm}[1]{\mbox{\boldmath$#1$}}
\newcommand{\be}{\begin{equation}}
\newcommand{\ee}{\end{equation}}
\newcommand{\bea}{\begin{eqnarray}}
\newcommand{\eea}{\end{eqnarray}}
\newcommand{\non}{\nonumber}
\newtheorem{df}{Definition}
\newtheorem{th1}[df]{Theorem}
\newtheorem{lem}[df]{Lemma}
\newtheorem{conj}[df]{Conjecture}
\newtheorem{prop}[df]{Proposition}
\newtheorem{cor}[df]{Corollary}
\newtheorem{lem2}{Lemma}[section]
\newtheorem{prop2}{Proposition}[section]
\newtheorem{cor2}{Corollary}[section]

\newcommand{\al}{\alpha}

\begin{document}

\title{Irreducibility criterion for a finite-dimensional 
highest weight representation of the $sl_2$ loop algebra 
and the dimensions of reducible representations}

\author{Tetsuo Deguchi
\footnote{e-mail deguchi@phys.ocha.ac.jp}}

\date{}

\maketitle

\begin{center}  
Department of Physics, Ochanomizu University, \\
2-1-1 Ohtsuka, Bunkyo-ku,Tokyo 112-8610, Japan
\end{center} 
\abstract{
We present a necessary and sufficient condition for a finite-dimensional highest weight representation of the $sl_2$ loop algebra to be irreducible. In particular, for a highest weight representation with degenerate parameters of the highest weight, we can explicitly determine whether it is irreducible or not.  We also present an algorithm for constructing finite-dimensional highest weight representations with a given highest weight. We give a conjecture that all the highest weight representations with the same highest weight can be constructed by the algorithm. For some examples we show the conjecture explicitly. The result should be useful in analyzing the spectra of integrable lattice models related to roots of unity representations of quantum groups, in particular, the spectral degeneracy of the XXZ spin chain at roots of unity associated with the $sl_2$ loop algebra. }

%
\newpage 
%
\section{Introduction}

Symmetry operators play a central role in the spectra of quantum systems.  
 Recently, it was shown that the XXZ spin chain at roots of unity  
commutes with the $sl_2$ loop algebra \cite{DFM}, and there exist  
 large spectral degeneracies associated with the symmetry 
 \cite{FM1,FM2,Odyssey,HWT,RAQIS05,NANKAI05}. 
The Hamiltonian of the XXZ spin chain 
under the periodic boundary conditions is given by   
\begin{equation} 
H_{XXZ} =  {\frac 1 2} \sum_{j=1}^{L} \left(\sigma_j^X \sigma_{j+1}^X +
 \sigma_j^Y \sigma_{j+1}^Y + \Delta \sigma_j^Z \sigma_{j+1}^Z  \right) \, . 
\label{hxxz}
\end{equation}
Here the XXZ anisotropic coupling $\Delta$ 
is related to the $q$ parameter by  $\Delta= (q+q^{-1})/2$. 
The symmetry of the $sl_2$ loop algebra 
appears when $q$ is a root of unity. 
The symmetry also appears in the spectrum of 
the transfer matrix of the six-vertex model at roots of unity \cite{DFM}.  
We note that  
the XXZ Hamiltonian is derived from the logarithmic derivative 
of the transfer matrix of the six-vertex model.   

Let us introduce the $sl_2$ loop algebra, $U(L(sl_2))$ 
\cite{Drinfeld,Chari,Chari-P1}. The generators ${x}_k^{\pm}$ and 
${h}_k$ for $k \in {\bf Z}$ satisfy  
\begin{eqnarray} 
{[} {h}_j, {x}_{k}^{\pm} {]} & = & \pm 2 {x}_{j+k}^{\pm} \, , \quad 
{[} {x}_j^{+}, {x}_k^{-} {]} = {h}_{j+k} \, , \non \\ 
{[} {h}_j, {h}_{k} {]} & = & 0 \, , 
\quad {[} {x}_j^{\pm}, {x}_k^{\pm} {]} = 0 \, , 
\quad {\rm for} \, \, j, k 
\in {\bf Z} \, . 
\label{CDR}
\end{eqnarray}
In a representation of $U(L(sl_2))$, we call a vector $\Omega$ 
{\it highest weight} if  $\Omega$ 
is annihilated by generators ${x}_{k}^{+}$ 
for all integers $k$ and such that 
$\Omega$ is a simultaneous eigenvector of every generator 
${h}_k$ ($k\in {\bf Z}$)  
\cite{Chari,Chari-P1,Chari-P2}: 
\bea 
{x}_k^{+} \Omega &= & 0 \, , \quad {\rm for} \, \, k \in {\bf Z} \, , 
\label{eq:annihilation} \\ 
{h}_{k} \Omega & = & {d}_k \Omega \, , 
\quad {\rm for} \, \, k \in {\bf Z} \, . 
\label{eq:Cartan}
\eea
We call the set of eigenvalues $d_k$  the highest weight of $\Omega$. 
The representation generated by a highest weight vector $\Omega$ 
is called the highest weight representation of $\Omega$. 
In the paper we assume that $\Omega$ generates a finite-dimensional 
representation.

It is easy to show that the weight $d_0$ is given by a nonnegative integer. 
We denote it by $r$. 
We shall show that $\Omega$ is a simultaneous eigenvector of  
operators $(x_0^{+})^{n}(x_1^{-})^{n}$:   
\be 
(x_0^{+})^{n}(x_1^{-})^{n} \Omega = (n!)^2 \, \lambda_n \Omega \, , \quad 
{\rm for} \quad n=1, 2, \ldots, r \, . 
\ee
In terms of $\lambda_n$'s,  
we define a polynomial ${\cal P}_{\lambda}(u)$ as follows: 
\be 
{\cal P}_{\lambda}(u) = \sum_{k=0}^{r} (-1)^k \, \lambda_k \, u^{k} \, .  
\ee
Here $\lambda$ denotes a sequence of $\lambda_n$'s, i.e.  
$\lambda=(\lambda_1, \lambda_2, \ldots, \lambda_r)$. 
We call ${\cal P}_{\lambda}(u)$ 
 the highest weight polynomial for the highest weight $d_k$.  
The highest weight polynomial generalizes the Drinfeld polynomial. 
In fact,  every finite-dimensional highest weight 
 representation has its highest weight polynomial. 
If the representation is irreducible, the highest weight polynomial 
 ${\cal P}_{\lambda}(u)$ is nothing but 
 the Drinfeld polynomial \cite{Chari-P1}. 
Let us factorize polynomial ${\cal P}_{\lambda}(u)$ as follows: 
\be 
{\cal P}_{\lambda}(u) = (1 - {\hat a}_1 u) \cdots (1 - {\hat a}_r u) \, . 
\ee 
We call parameters ${\hat a}_1, {\hat a}_2, \ldots, 
{\hat a}_r$, the {\it highest weight parameters} of $\Omega$. 
In terms of highest weight parameters ${\hat a}_j$,   
the highest weight $d_k$ of $\Omega$ are expressed as  
\be 
d_k = \sum_{j=1}^{r} {\hat a}_j^k  \quad {\rm for} \, \, k \in {\bf Z} \, ,  
\ee
and the eigenvalues $\lambda_k$ are expressed as    
\be 
\lambda_k = \sum_{1 \le i_1 < i_2 < \cdots < i_k \le r} 
{\hat a}_{i_1} {\hat a}_{i_2} \cdots {\hat a}_{i_k} \, .  
\ee   

Let us discuss how to evaluate degenerate multiplicities 
in the spectrum of some model that has the $sl_2$ loop algebra symmetry. 
They are given by the dimensions of some 
 finite-dimensional representations of the $sl_2$ loop algebra.    
Here we have a conjecture that every finite-dimensional representation 
is decomposed into a collection of finite-dimensional 
highest weight representations.  
Thus, the degenerate multiplicities should  be evaluated 
essentially in terms of the dimensions of 
corresponding finite-dimensional highest weight representations.     
The dimensions of irreducible representations of 
 $U(L(sl_2))$ and those of $U_q(L(sl_2))$  
are known \cite{Chari,Chari-P1,Chari-P2}. 
Furthermore, it was shown by Chari and Pressley \cite{Chari-P3} 
that corresponding to 
each irreducible finite-dimensional representation 
with highest weight parameters, ${\hat a}_1, {\hat a}_2, \ldots, 
{\hat a}_r$, there exists a unique finite-dimensional 
highest weight module $W$ with the highest weight parameters 
${\hat a}_j$ for $j=1,2, \ldots, r$,  
such that any finite-dimensional highest weight module $V$ 
with the same highest weight parameters ${\hat a}_j$ ($j = 1, 2, \ldots, r$),  
is a quotient of $W$. We call $W$ the Weyl module 
of the highest weight parameters ${\hat a}_j$.  
It has also been shown \cite{Chari-P3} that a Weyl module is irreducible 
if and only if the highest weight parameters ${\hat a}_j$ are distinct.  
However, if some of the highest weight parameters are degenerate, 
it is not trivial whether the representation generated by $\Omega$ 
is irreducible or not. 

In the paper,  we prove a necessary and sufficient condition   
for a finite-dimensional highest weight representation 
to be irreducible. 
Suppose that a highest weight vector $\Omega$ with highest weight 
parameters  ${\hat a}_j$ for $j = 1, 2, \ldots, r$,  
generates a finite-dimensional representation $U \Omega$.  
We also assume that the highest weight parameters 
${\hat a}_j$ for $j = 1, 2, \ldots, r$ are given by 
a set of distinct parameters  
$a_j$ with multiplicities $m_j$ for $1 \le j \le s$.  Then, 
we shall show that  $U \Omega$ is irreducible 
if and only if the following condition holds:  
\be 
\sum_{j=0}^{s} (-1)^{s-j} \mu_{s-j} { x}_{j}^{-} \Omega = 0  \, ,  
\label{eq:cond}
\ee
where coefficients $\mu_{k}$ $(k=1, 2, \ldots, s)$ are given by 
\be 
\mu_{k} = \sum_{1 \le i_1 < i_2 < \cdots < i_k \le s} 
a_{i_1} a_{i_2} \cdots a_{i_k}  
\, . 
\ee
If $U \Omega$ is irreducible, the dimensionality is given by  
\be 
{\rm dim} \, U \Omega= \prod_{j=1}^{s} (m_j+1) . 
\ee
Here we note that if the highest weight parameters ${\hat a}_j$ 
are distinct (i.e. $s=r$), 
condition (\ref{eq:cond}) holds trivially. 
Furthermore, we introduce an algorithm by which  
we can construct practically all finite-dimensional 
highest weight modules with a given set of 
highest weight parameters ${\hat a}_j$. 
It is a conjecture that 
all such representations are constructed by the algorithm.   
Here we note that the algorithm is not complete in the sense that   
we employ  some conjectured relations among products of generators 
acting on $\Omega$, by which we exclude some redundant 
quotients of submodules.  
For some simple cases, however,  we explicitly calculate  
dimensions of all possible reducible highest weight  
representations that have the same given set of 
highest weight parameters ${\hat a}_j$.

As an illustration, let us consider the case of $r=3$ where two of the 
 three highest weight parameters are degenerate, i.e.  
 $({\hat a}_1, {\hat a}_2, {\hat a}_3)=(a_1, a_1, a_2)$. 
We have 
$$ 
{\cal P}_{\lambda}(u) = ( 1- a_1 u)^2( 1- a_2 u) = 1 - (2 a_1 + a_2) u + 
(a_1^2 + 2 a_1 a_2) u^2 - a_1^2 a_2 u^3 \, . 
$$
 For any highest weight vector $\Omega$ 
with the same highest weight  we have 
$$ 
\left( x_3^{-} - (2 a_1 + a_2) x_2^{-} + (a_1^2 + 2 a_1 a_2) x_1^{-} 
- a_1^2 a_2 x_0^{-} \right) \Omega = 0 \, .  
$$
However, if  $\Omega$ satisfies the following relation: 
$$ 
\left( x_2^{-} - (a_1 + a_2) x_1^{-} + a_1 a_2 x_0^{-} \right) \Omega \, 
= 0 \, ,   
$$
then it generates an irreducible representation.

The highest weight polynomial should be useful for physical applications. 
Let us recall that every finite-dimensional highest weight representation 
has a unique highest weight polynomial ${\cal P}_{\lambda}(u)$, 
while it has the Drinfeld polynomial only if it is irreducible.  
We also recall that the degenerate eigenspaces of some physical system 
that has the $sl_2$ loop algebra symmetry 
should be given by collections of 
finite-dimensional highest weight representations. 
However, it is not certain  
whether they are irreducible or not.  
We therefore introduce the highest weight polynomial, by which 
we can investigate highest weight representations 
that are not necessarily irreducible.

The general criterion of irreducibility 
should be useful for studying the spectra of some integrable models 
associated with roots of unity representations of the quantum groups, 
in particular, the spectral degeneracy of 
the XXZ spin chain and the six-vertex model at roots of unity \cite{HWT}.  
Recall that in order to derive the degenerate multiplicities 
of the $sl_2$ loop algebra symmetry rigorously, one has to calculate  
the dimensions of highest weight representations generated by 
the corresponding Bethe vectors \cite{Odyssey,HWT}.  
However, it is not trivial whether they are 
irreducible or not. In fact, there exists 
such a Bethe vector that 
is highest weight and generates a reducible representation \cite{HWT}.  
Furthermore, it has not been discussed how to derive 
 the dimensions of reducible highest weight representations 
in the general case. Thus, through the irreducibility criterion 
and the algorithm for constructing practically all reducible representations 
with the same highest weight, we can evaluate  
 the degenerate multiplicities systematically. 
It should be remarked 
that the degenerate eigenvectors of the $sl_2$ loop algebra 
can also be discussed in terms of Bethe vectors 
through some limiting procedure \cite{Baxter}.  
However, such limiting procedure 
is not always straightforward \cite{FM1,FM2}, and 
it does not give a systematic method. 
We also note that the irreducibility criterion (\ref{eq:cond}) 
has been announced in Refs. 
\cite{HWT,RAQIS05,NANKAI05} without an explicit proof for the general case.

There are several important viewpoints associated with the 
$sl_2$ loop algebra symmetry of the six-vertex model at roots of unity.  
(i) Roots of unity representations of 
the quantum groups  have many subtle and interesting properties 
\cite{Arnaudon,DeConcini,Modular,Lusztig}. 
They have several connections to the six-vertex model 
 and the XXZ spin chain at roots of unity \cite{Rittenberg,Pasquier}. 
Roots of unity representations are 
also related to the chiral Potts model \cite{Kyoto,Tarasov}.  
(ii) It has been shown for the $sl_2$ loop algebra that 
 every irreducible representation is 
given by a tensor product of evaluation modules with distinct 
evaluation parameters \cite{Chari}. However, it has not been discussed  
how to determine whether  
a given highest weight representation is irreducible or not. 
For $U_q(L(sl_2))$, the irreducibility criterion for 
tensor products of evaluation modules  has been shown 
in Ref. \cite{Chari-P1}. For Yangians, 
an irreducibility criterion for 
tensor products of Yangian evaluation modules has been derived  
in Ref. \cite{Molev}. 
(iii) The Bethe ansatz equations of integrable lattice models 
associated with Yangians and reflection 
algebras are discussed \cite{Arnaudon2}. Here, the Drinfeld polynomials 
play an important role. 
(iv) Quantum groups at roots of unity have cyclic 
representations which have no highest weight vector \cite{Arnaudon}. 
It has been conjectured that some highest weight representation 
of the $sl_2$ loop algebra is closely connected to the Onsager 
algebra symmetry of the chiral Potts model 
at the superintegrable point \cite{Nishino}. 
Recall that the chiral Potts model is related to  
the cyclic representations of quantum groups \cite{Kyoto}.    
Thus, the result of the present paper might be useful 
also for future researches in integrable systems associated with  
roots of unity representations.

The $sl_2$ loop algebra symmetry of the XXZ spin chain at roots of unity 
should be interesting also in the spectral analysis 
of quantum Hamiltonians. In the spectral flow 
of the XXZ spin chain with respect to $\Delta$, 
level crossings with exponentially large degenerate multiplicities 
appear at some discrete values of $\Delta$.  
According to a theorem by von Neumann and Wigner,  it is far 
more likely to have spectral degeneracies if one or more symmetries exist 
than if no symmetries exist \cite{Neumann-Wigner}. 
In association with it, the noncrossing rule is stated as follows: 
in the spectrum of a Hamiltonian that depends on a real parameter,   
levels of the same symmetry never cross each other as the parameter 
varies, where all levels are classified by symmetry quantum numbers. 
Here we call an operator commuting with the Hamiltonian a symmetry operator 
if it is independent of the model parameter \cite{Heilmann-Lieb}.  
However, the spectrum of a Hamiltonian can have degeneracies. 
Novel counterexamples to the noncrossing rule were discussed   
in the spectrum of the one-dimensional Hubbard Hamiltonian 
\cite{Heilmann-Lieb}.   

The paper is organized as follows.  
In \S 2, we introduce generators of the $sl_2$ loop algebra with 
parameters. In \S 3, we discuss sectors of highest weight representations.  
We show that every finite-dimensional 
highest weight representation has 
 nonzero and finite highest weight parameters.  
In \S 4, we define highest weight parameters and highest weight polynomials, 
explicitly. 
In \S 5, we prove the irreducibility criterion. In \S 6, we formulate 
an algorithm by which we can  construct practically all  
reducible or irreducible 
highest weight representations that have the same highest weight. 
 For two simple cases, we derive dimensions of 
all reducible highest weight modules that have the same highest weight, 
explicitly.

Throughout  the paper we assume that  
 $\Omega$ is a non-zero highest weight vector with highest weight 
$d_k$ in a finite-dimensional representation of $U(L(sl_2))$. 
Thus, it generates a finite-dimensional representation,  
which we denote by $U \Omega$. 
We  denote the highest weight $d_0$ by $r$, i.e. ${h}_0 \Omega= r \Omega$. 
As shown in \S 3, $r$ is given by a nonnegative integer, and 
it is equal to the number of highest weight parameters ${\hat a}_j$ 
of $\Omega$.

%
\section{Loop algebra generators with parameters }
%

Let ${\bm \alpha}$ denote a finite sequence of complex parameters such as 
${\bm {\alpha}} =(\alpha_1, \alpha_2, \ldots, \alpha_n)$. 
We define generators with $n$ parameters, $x_m^{\pm}({\bm \al})$ and 
$h_m({\bm \al})$, as follows \cite{RAQIS05,NANKAI05}: 
\begin{eqnarray} 
x_{m}^{\pm}({\bm \al}) & = & 
\sum_{k=0}^{n} (-1)^k x_{m-k}^{\pm} 
\sum_{\{ i_1, \ldots, i_k \} \subset \{1, \ldots, n \}}  
\alpha_{i_1} \alpha_{i_2} \cdots \alpha_{i_k} \, , 
\non \\
h_{m}({\bm \al}) & = & 
\sum_{k=0}^{n} (-1)^k h_{m-k} 
\sum_{\{ i_1, \ldots, i_k \} \subset \{1, \ldots, n \}}  
\alpha_{i_1} \alpha_{i_2} \cdots \alpha_{i_k} \, . 
\label{eq:def-alpha}
\end{eqnarray}
Let ${\bm \al}$ and ${\bm \beta}$ 
be arbitrary sequences of $n$ and $p$ parameters, respectively.  
In terms of generators with parameters 
we express the defining relations of the $sl_2$ loop algebra as follows:   
\begin{equation} 
[x_{\ell}^{+}({\bm \al}), x_m^{-}({\bm \beta})] 
= h_{\ell+m}({\bm {\al \beta}}) \, , \quad 
[h_{\ell}({\bm \al}), x^{\pm}_{m}({\bm \beta})] 
= \pm 2  x_{\ell+m}^{\pm}({\bm {\al \beta}}) \, . 
\label{eq:dfr-AB}
\end{equation}
Here the symbol ${\bm {\al \beta}}$ 
denotes the composite sequence of ${\bm \al}$ and ${\bm \beta}$: 
\be 
{\bm {\al \beta}} 
= (\alpha_1, \alpha_2, \ldots, \alpha_{n}, 
\beta_1, \beta_2, \ldots, \beta_{p}). 
\ee
Using  relations (\ref{eq:dfr-AB}), 
we can show the following relations for $t \in {\bf Z}_{\ge 0}$: 
\begin{eqnarray} 
{[}  (x_m^{+}({\bm \al}))^{(t)},  x_{\ell}^{-}({\bm \beta}) {]} 
& = & (x_m^{+}({\bm \al}) )^{(t-1)} 
 h_{\ell+m}({\bm \al} {\bm \beta}) 
 +  x_{\ell+2m}^{+}({\bm \al} {\bm \al}{\bm \beta}) 
 (x_{m}^{+}({\bm \al}))^{(t-2)} \, , \non \\ 
 {[} x_{\ell}^{+}({\bm \al}), (x_m^{-}({\bm \beta}))^{(t)} {]} 
 & = & (x_m^{-}({\bm \beta}) )^{(t-1)} 
 h_{\ell+m}({\bm \al} {\bm \beta}) 
 - x_{\ell+2m}^{-}({\bm \al} {\bm \beta} {\bm \beta}) 
 (x_{m}^{-}({\bm \beta}))^{(t-2)} \, , \non \\ 
 {[} h_{\ell}({\bm \al}), (x_m^{\pm}({\bm \beta}))^{(t)} {]} 
 & = & \pm 2 (x_m^{\pm}({\bm \beta}))^{(t-1)} 
 x_{\ell+m}^{\pm}({\bm \al} {\bm \beta}) \, . \label{eq:AB}
\end{eqnarray}
Here the symbol $(X)^{(n)}$ denotes the $n$th power of operator $X$ 
divided by the $n$ factorial, i.e. $(X)^{(n)} = X/n!$.

For a given sequence of $m$ parameters, 
${\bm \al}=(\al_1, \al_2, \cdots, \al_m)$,  
let us denote by ${\bm \al}_{j}$ the sequence of parameters of 
${\bm \al}$  other than $\al_j$:  
\be 
{\bm \al}_{j} = 
(\al_1, \ldots, \al_{j-1}, \al_{j+1}, \ldots, \al_m ) . 
\ee  
Here we assume that parameters $\alpha_j$ take distinct values for 
$j=1, 2, \ldots, m$.  
We now introduce the following symbol: 
\begin{equation} 
\rho_{j}^{\pm}({\bm \al}; m) = x_{m-1}^{\pm}({\bm \al}_j) \, \quad 
{\rm for } \quad j = 1, 2, \ldots, m . 
\end{equation} 
The generators $x_{j}^{\pm}$ for $j=0, 1, \ldots, m-1$,  
are expressed as linear combinations of 
$\rho_j^{\pm}({\bm \alpha}; m)$. 
Let us introduce a symbol $\al_{kj}$ by $\al_{kj}=\al_k - \al_j$.  
It is easy to show the following: 
\be 
(-1)^{n-1} \sum_{j=1}^{n} {\frac {\rho_j^{\pm}({\bm \alpha}; m)} 
{\prod_{k=1; k \ne j}^{n} \al_{kj}} } 
= x_{m-n}^{\pm}(\al_{n+1}, \ldots, \al_{m})  \quad (1 \le n \le m)  \, . 
\label{eq:rho-x}
\ee   
It can be shown by making use of eqs. (\ref{eq:rho-x})  
that $x_{k}^{\pm}$ ($0 \le k \le m-1$) 
are expressed in terms of linear combinations 
of $\rho_j^{\pm}({\bm \alpha}; m)$ with $1 \le j \le m$.

%
\section{Sectors of a highest weight representation}
%

Let us briefly introduce elementary representation theory of $sl_2$ 
as follows.  
\begin{lem}
Let $e, f$ and $h$ be standard generators of $sl_2$. 
If $u$ is a non-zero vector in a finite-dimensional representation of 
$sl_2$ such that $eu=0$ and $h u = r u$, then we have the following: 
(i) $r$ is a non-negative integer;   
(ii) $U(sl_2) u$ is an irreducible $(r+1)$-dimensional representation;  
(iii) $f^{r} u \ne 0$ and $f^{r+1} u = 0$.  
\label{lem:sl2}
\end{lem}

We now consider the $sl_2$-subalgebra ${\cal U}_k$ generated by 
$x_{-k}^{+}$, $x_k^{-}$ and $h_0$ for an integer $k$. 
Here we recall that  $\Omega$ is a non-zero highest weight vector 
with highest weight $d_k$ 
in a finite-dimensional representation  of $U(L(sl_2))$. 
Applying the Poincar{\' e}-Birkhoff-Witt theorem \cite{Jacobsen} 
to $U(L(sl_2))$,  we can show that  in $U \Omega$ 
 eigenvalues of $h_0$ are given by integers. We call them weights. 
It also follows that $U \Omega$ is given by 
the direct sum of subspaces of weights, 
i.e. sectors with respect to eigenvalues of $h_0$.  
Applying lemma \ref{lem:sl2} to ${\cal U}_k$ we have the following.  
\begin{cor}
The highest weight $d_0$ is given by a non-negative integer, $r$, and we have  
$(x_k^{-})^r \Omega \ne 0$ and $(x_k^{-})^{r+1} \Omega = 0$ 
for $k \in {\bf Z}$. 
\label{cor:UO}
\end{cor} 

\begin{prop} 
The subspace of weight $-r$ of $U \Omega$ is one-dimensional. 
\label{prop:1-dim} 
\end{prop}
We shall discuss an explicit proof of 
proposition \ref{prop:1-dim} in Appendix A. 

\begin{lem} 
$\Omega$ is a simultaneous eigenvector of $(x_0^{+})^{(n)} (x_1^{-})^{(n)}$:   
\be 
(x_0^{+})^{(j)} (x_1^{-})^{(j)} \Omega = \lambda_j \Omega \, , \quad 
\mbox{\rm for} \quad j=1, 2, \ldots, r \, . 
\label{eq:01}
\ee
Here $\lambda_j$ are eigenvalues. 
\end{lem} 
\begin{proof} 
Applying the Poincar{\' e}-Birkhoff-Witt theorem \cite{Jacobsen} 
to $U(L(sl_2))$, we derive that  the subspace 
of weight $r$ in $U \Omega$ is one-dimensional. 
Here we note that  $(x_0^{+})^{(j)} (x_1^{-})^{(j)} \Omega$ is 
in the subspace of weight $r$ in $U \Omega$. 
Therefore, 
$(x_0^{+})^{(j)} (x_1^{-})^{(j)} \Omega$ is proportional to the 
basis vector $\Omega$, and hence we have (\ref{eq:01}). 
\end{proof} 

\begin{prop} 
Eigenvalue $\lambda_r$  is non-zero. 
Here we recall 
$({  x}_0^{+})^{(r)} ({  x}_1^{-})^{(r)} \, \Omega 
= \lambda_r \, \Omega$. 
\label{prop:nonzero-roots} 
\end{prop}
\begin{proof}
Recall corollary \ref{cor:UO} that 
$({  x}_1^{-})^{r} \Omega \ne 0 $ and $({  x}_0^{-})^{r} \Omega \ne 0$.  
It follows from proposition \ref{prop:1-dim} that  
they are linearly dependent, i.e. we have  
$({  x}_1^{-})^{r} \Omega= A_1 ({  x}_0^{-})^{r} \Omega$  
with a nonzero constant $A_1$.    
The eigenvalue $\lambda_r$ is given by $A_1$ as follows:  
\be 
\lambda_r (r!)^2 \, \Omega 
= ({  x}_0^{+})^{r} \, ({  x}_1^{-})^{r} \Omega = 
A_1 ({  x}_0^{+})^{r} ({  x}_0^{-})^{r} \Omega
= A_1 \, (r!)^2 \, \Omega  
\ee
We thus obtain $\lambda_r = A_1$ and $\lambda_r \ne 0$. 
\end{proof}

%
\section{Highest weight polynomials}
%

\subsection{Parameters expressing the highest weight}

We now introduce parameters expressing the highest weight 
of $\Omega$.  
Let $\lambda=(\lambda_1, \ldots, \lambda_r)$ 
denote the sequence of eigenvalues $\lambda_k$ which are defined 
in eq. (\ref{eq:01}).   
We define a polynomial $P_{\lambda}(u)$ by the following relation 
\cite{Jimbo-summer}:  
\be 
P_{\lambda}(u) =\sum_{k=0}^{r} \lambda_k (-u)^k  \, .   
\label{eq:DrinfeldP}
\ee
We call it the {\it highest weight polynomial} of $\Omega$.

Let us factorize polynomial $P_{\lambda}(u)$ as follows 
\be 
P_{\lambda}(u) = \prod_{k=1}^{s} (1 - a_k u)^{m_k} \, ,   
\label{eq:factor}
\ee 
where $a_1, a_2, \ldots, a_s$ are distinct, and their  
 multiplicities are given by  $m_1, m_2, \ldots, m_s$, respectively.   
We denote by ${\bm a}$ the sequence of $s$ 
parameters $a_j$: 
\be 
{\bm a}=(a_1, a_2, \ldots, a_s). 
\ee
Here we note that $r$ is equal to 
the sum of multiplicities $m_j$: $r=m_1 + \cdots + m_s$. 
We define parameters ${\hat a}_i$ for $i=1, 2, \ldots, r$, as follows.  
\be 
{\hat a}_i = a_k \quad {\rm if } \, \, m_1+ m_2 + \cdots + m_{k-1} < i \le  
m_1+ \cdots + m_{k-1} + m_{k} \, . 
\label{eq:hat-a}
\ee
Then, the set $\{ {\hat a}_j \, | j =1, 2, \ldots, r \}$ corresponds to the 
set of parameters $a_j$ with multiplicities $m_j$ for 
 $j=1, 2, \ldots, s$. We denote by ${\hat {\bm a}}$ 
 the sequence of $r$ parameters ${\hat a}_i$: 
\be  
 {\hat {\bm a}}=({\hat a}_1, {\hat a}_2, \ldots, {\hat a}_r) .      
\ee
We call parameters ${\hat a}_i$ 
the {\it highest weight parameters} of $\Omega$. 
It follows from the definition of highest weight polynomial 
${\cal P}_{\lambda}(u)$ given by (\ref{eq:DrinfeldP}) 
and that of highest weight parameters (\ref{eq:factor}) that 
 we have   
\be 
\lambda_{n} = \sum_{1 \le j_1 < \cdots < j_n \le r} 
{\hat a}_{j_1} \cdots {\hat a}_{j_n} \, . 
\label{eq:lambda}
\ee

\begin{prop}
The roots of highest weight polynomial $P_{\lambda}(u)$ 
are non-zero, and the degree is given by $r$.  
\end{prop}
\begin{proof} 
Recall proposition \ref{prop:nonzero-roots} that $\lambda_r \ne 0$. 
We note that $\lambda_r = \prod_{j=1}^{r} {\hat a}_j =  
\prod_{j=1}^{s} a_j^{m_j}$. Therefore, $a_j$ are non-zero 
for $j=1, 2, \ldots, s$.   
\end{proof}

\subsection{Recursive lemmas}

Let $a$ be an arbitrary complex number. 
We denote by $(a)^{n}$ 
the sequence of parameter $a$ with multiplicity $n$, 
i.e. $(a)^{n}= (a, a, \ldots, a)$.  
For $n=1$ we write 
$x_n^{\pm}((a)^{1}) = x_n^{\pm} - a \, x_{n-1}^{\pm}$ 
simply as $x_n^{\pm}(a)$. 
We also introduce the following: 
 \be 
\lambda_{n}(a) = \sum_{1 \le j_1 < \cdots < j_n \le r} 
({\hat a}_{j_1} -a) \cdots ({\hat a}_{j_n}-a) \, . 
\label{eq:lambda-a}
\ee

\begin{lem} 
For a given integer $\ell$, we have 
\bea 
({  x}_{\ell}^{+})^{(n)} ({  x}_{1-\ell}^{-}(a))^{(n+1)} 
& = & {  x}_{1-\ell}^{-}(a) \,  
({  x}_{\ell}^{+})^{(n)} ({  x}_{1-\ell}^{-}(a))^{(n)} 
+ {\frac 1 2} \, {[} {  h}_{1}(a), ({  x}_{\ell}^{+})^{(n-1)} 
({  x}_{1-\ell}^{-}(a))^{(n)} {]} 
\non \\
& & \quad - ({  x}_{\ell}^{+})^{(n-1)} 
({  x}_{1-\ell}^{-}(a))^{(n+1)} {  x}_{\ell}^{+} \, , \quad 
 {\rm for} \,\,  n \in {\bf Z}_{\ge 0} \, .  
\label{eq:ind-ell} 
\eea
\label{lem:ind-ell}
\end{lem}
\begin{proof} 
We first show the following relations by induction on $n$: 
\bea 
{[} {  h}_1(a), ({  x}_{\ell}^{+})^{(n)} {]} & = & 
2 {  x}_{\ell+1}^{+}(a) ({  x}_{\ell}^{+})^{(n-1)} \, , \non \\ 
{[} ({  x}_{\ell}^{+})^{(n)}, {  x}_{1-\ell}^{-}(a) {]} & = & 
({  x}_{\ell}^{+})^{(n-1)} {  h}_1(a) 
+ {  x}_{\ell+1}^{+}(a) ({  x}_{\ell}^{+})^{(n-2)} \, , \non \\ 
{[} {  h}_1(a), ({  x}_{1-\ell}^{-}(a))^{(n)} {]} & = & 
(-2) \, {  x}_{2-\ell}^{-}((a)^2) 
({  x}_{1-\ell}^{-}(a))^{(n-1)} \, , \non \\  
{[} {  x}_{\ell}^{+}, ({  x}_{1-\ell}^{-}(a))^{(n)} {]} & = & 
({  x}_{1-\ell}^{-}(a))^{(n-1)} {  h}_1(a) 
- {  x}_{2-\ell}^{-}((a)^2) ({  x}_{1-\ell}^{-}(a))^{(n-2)} \, .  
\label{four-rec}
\eea
Making use of relations (\ref{four-rec}) we can show  relation 
(\ref{eq:ind-ell}). Some details will be shown in Appendix B.  
\end{proof}

In the case of $a=0$ and $\ell=0$ 
the relation (\ref{eq:ind-ell}) has been shown for the case of 
$U_q(L(sl(2)))$ \cite{Chari-P1}.

Let $U$ denote the $sl_2$ loop algebra  $U(L(sl_2))$. 
For a given integer $\ell$, 
let $U({\cal B}_{\ell})$ be the subalgebra of $U(L(sl_2))$ 
generated by ${ h}_k, {x}_{\ell+ k}^{+}$  
and ${x}_{-\ell+1+k}^{-}$ for $k \in {\bm Z}_{\ge 0}$. 
We denote by ${\cal B}_{\ell}^{+}$ such a subalgebra of $U({\cal B}_{\ell})$ 
that is generated by ${x}_{\ell + k}^{+}$ for $k \in {\bm Z}_{\ge 0}$. 
\begin{lem} 
For a given integer $\ell$ we have  
the following recursive relations for $n \in {\bm Z}$:    
\bea
&({\rm A}_n):&  ({  x}_{\ell}^{+})^{(n-1)} ({  x}_{1-\ell}^{-}(a))^{(n)}  
 =  \sum_{k=1}^{n} (-1)^{k-1} {  x}_{k-\ell}^{-}((a)^k) 
({  x}_{\ell}^{+})^{(n-k)} 
({  x}_{1-\ell}^{-}(a))^{(n-k)} \, \mbox{\rm mod} \, 
U({\cal B}_{\ell}){\cal B}_{\ell}^{+},   \non \\
&({\rm B}_n):&   
({  x}_{\ell}^{+})^{(n)} ({  x}_{1-\ell}^{-}(a))^{(n)}  
 =  {\frac 1 n} \, \sum_{k=1}^{n} (-1)^{k-1} {  h}_{k}((a)^k) 
 ({  x}_{\ell}^{+})^{(n-k)} 
 ({  x}_{1-\ell}^{-}(a))^{(n-k)} \, \mbox{\rm mod} \,
  U({\cal B}_{\ell}) {\cal B}_{\ell}^{+},  \non \\
&({\rm C}_n):&  
  {[} {  h}_j(a),  ({  x}_{\ell}^{+})^{(m)} 
({  x}_{1-\ell}^{-}(a))^{(m)} {]}  = 0 \,  
\mbox{\rm mod} \, U({\cal B}_{\ell}) {\cal B}_{\ell}^{+} \, \quad 
\mbox{\rm for} \, \, m \le n \, \,  \mbox{\rm and} \, \,  
j \in {\bm Z} \, . \non
\eea
\label{lem:ABC}
\end{lem} 
\begin{proof} 
We now show relations $({\rm A}_n)$, $({\rm B}_n)$ and 
 $({\rm C}_n)$, 
inductively on $n$ as follows:  
We first show $({\rm A}_1)$, $({\rm A}_2)$, $({\rm B}_1)$ and $({\rm C}_1)$, 
directly. Then, 
relation $({\rm A}_n)$ is derived from $({\rm A}_{n-1})$ 
and $({\rm C}_{n-2})$. 
Here we make use of formula (\ref{eq:ind-ell}).  
Relation $({\rm B}_n)$ is derived from $({\rm A}_n)$, $({\rm C}_{n-1})$ 
and $({\rm B}_{m})$ for $m \le n-1$. 
We multiply both hand sides of $({\rm A}_n)$ by ${x}_{\ell}^{+}$ 
from the left. We show 
$x_{\ell}^{+} (x_{\ell}^{+})^{(m)} (x_{1-\ell}^{-}(a))^{(m)} 
\in U({\cal B}_{\ell}) {\cal B}_{\ell}^{+}$ for $m \le n-1$ 
by induction on $m$. Here we make use of $({\rm B}_{m})$ for $m \le n-1$ 
and ${\rm C}_{n-1}$. 
Finally, 
 $({\rm C}_n)$ is derived from $({\rm B}_{n-1})$ and $({\rm C}_{n-1})$.  
Thus, the cycle of induction process, 
$({\rm A}_n)$, $({\rm B}_n)$ and $({\rm C}_n)$, 
is closed. 
\end{proof}

Applying relations $({\rm B}_n)$ of lemma \ref{lem:ABC}, 
we shall prove an irreducibility criterion in \S 5. 

\subsection{Reduction relations}

%
%
%
%
Let us define the elementary symmetric polynomials $p_m$ 
in $x_1, \ldots, x_r$, as follows:  
\be 
p_m = \sum_{1 \le i_1 < \cdots < i_m \le r} x_{i_1} \cdots x_{i_m} \, .    
\ee 
We introduce symmetric polynomials $s_k$ by $s_k= \sum_{j=1}^{r} x_j^k$.   
Then, Newton's formulas are given as follows: 
\bea 
p_n & = & {\frac 1 n} \sum_{k=1}^{n} (-1)^{k-1} \, s_k \, p_{n-k} \, \quad \mbox{\rm for} \quad n \le r ,  \label{eq:Newton1} \\ 
s_{r+1+j} & = & \sum_{k=1}^{r} (-1)^{r-k} s_{k+j} \, p_{r+1-k} \quad \mbox{\rm for} \quad 
j \in {\bm Z}. \label{eq:Newton2}
\eea


We now show systematically that   
highest weights $d_k$ are expressed as the  
symmetric polynomials $s_k$ 
of highest weight parameters ${\hat a}_j$. 
Substituting $\ell=0$ and $a=0$ in $({\rm B}_n)$ of lemma \ref{lem:ABC} 
and making use of (\ref{eq:Cartan}),  
we have the following. 
\begin{cor}
\be 
\lambda_n = {\frac 1 n} \sum_{k=1}^{n} (-1)^{k-1} d_k \lambda_{n-k} \, 
 , \quad  \mbox{\rm for} \quad  n=1, 2, \ldots, r. 
\label{eq:dk-lambda}
\ee
\label{cor:dk-lambda}
\end{cor}

\begin{lem}
For any integer $\ell$ we have 
\be 
({  x}_{\ell}^{+})^{(n)} ({ x}_{1-\ell}^{-})^{(n)} 
\Omega = \lambda_n \Omega \, , \quad  \mbox{\rm for} \quad  
n = 1, 2, \ldots, r . \label{eq:xx-lambda}
\ee
\label{lem:lambda}
\end{lem} 
\begin{proof} 
Making use of $({\rm B}_n)$ with $a=0$, 
we show (\ref{eq:xx-lambda}) by induction on $n$. 
Here we also make use of (\ref{eq:dk-lambda}) and (\ref{eq:Cartan}).  
\end{proof}

\begin{prop}[Reduction relations]   
\bea 
{x}_{r+1-\ell}^{-} \, \Omega & = & \sum_{k=1}^{r} (-1)^{r-k} 
\lambda_{r+1-k} \, {x}^{-}_{k-\ell} \, \Omega \, ,  \quad {\rm for} \, \, 
\ell \in {\bf Z} \, ,  
\label{eq:rr-ell-x} \\ 
{d}_{r+1-\ell}  & = & \sum_{k=1}^{r} (-1)^{r-k} 
\lambda_{r+1-k} \, {d}_{k-\ell} \, , \quad {\rm for} \, \, 
\ell \in {\bf Z} \, .  
\label{eq:rr-ell-d}
\eea
\label{lem:rr-ell}
\end{prop} 
\begin{proof}
We derive  reduction relations (\ref{eq:rr-ell-x}) from 
$({\rm A}_{r+1})$ of lemma \ref{lem:ABC} with $a=0$ 
and lemma \ref{lem:lambda}. Applying $x_0^{+}$ to (\ref{eq:rr-ell-x}) 
from the left, 
we derive relations (\ref{eq:rr-ell-d}).   
\end{proof} 
\begin{prop} The highest weights $d_n$ are given by 
the following symmetric polynomials 
of highest weight parameters ${\hat a}_j$: 
\be 
d_n = \sum_{j=1}^{r} {\hat a}_j^{n}  \, , 
\quad \mbox{for} \, \, n \in {\bm Z} \, . 
\label{eq:dn} 
\ee
\end{prop} 
\begin{proof} 
Making use of (\ref{eq:dk-lambda})  
and Newton's formula (\ref{eq:Newton1}), 
we show (\ref{eq:dn}) by induction on $n$ for $n=1, 2, \ldots, r$. 
Then, we generalize (\ref{eq:dn}) to the case of arbitrary integers $n$ 
through Newton's formula (\ref{eq:Newton2}). 
\end{proof}

\begin{prop} 
Let ${\bm \alpha}=(\alpha_1, \ldots, \alpha_n)$ be a sequence of 
arbitrary complex parameters.   
For a given integer $m$  we have 
\be 
h_m({\bm \alpha}) \Omega = d_m({\bm \alpha}) \Omega \label{eq:h-alpha} \, , 
\ee
where $d_m({\bm \alpha})$ is given by 
\be 
d_m({\bm \alpha}) 
= \sum_{j=1}^{r} {\hat a}_j^{m-n} \prod_{i=1}^{n} ({\hat a}_j - \alpha_i) \, . 
\label{eq:d-alpha} 
\ee
In particular, if the set ${\bm \alpha}$ contains $a_1, a_2, \ldots$, 
and $a_s$,  we have 
\be
h_m({\bm \alpha}) \Omega  = 0 \, . 
\ee 
\label{prop:hmalpha} 
\end{prop} 
\begin{proof}
Through
(\ref{eq:rr-ell-d}) and (\ref{eq:Newton2}), 
we show (\ref{eq:dn}) for any integer $n$.  
Substituting relations (\ref{eq:dn}) for $n \in {\bm Z}$ 
into (\ref{eq:def-alpha}), we obtain (\ref{eq:h-alpha}).  
\end{proof}

Reduction relations (\ref{eq:rr-ell-x}) are expressed as follows: 
\be 
x_{r+ 1-\ell}^{-} ({\hat {\bm a}}) \Omega = 0 \quad 
\mbox{for} \, \, \ell \in {\bm Z }\, . 
\label{eq:red-all}
\ee
Reduction relations (\ref{eq:red-all}) are fundamental when 
we construct a reducible or irreducible highest weight representation.  
Here we note that relations (\ref{eq:rr-ell-x}) and 
(\ref{eq:red-all}) play a similar role 
as the characteristic equations of matrices.

\begin{lem} Let $\ell$ be an integer. 
If $x_{\ell}^{-}(A) \Omega=0$ for a sequence of parameters $A$,  then 
for any sequence of parameters $B$ that contains $A$ as a set, we have 
\be 
x_{m}^{-}(B) \, \Omega = 0 \quad \mbox{\rm for} \quad m \in {\bm Z} \, . 
\ee  
\label{lem:vanishAB}
\end{lem}  
\begin{proof} 
Let us denote by $C$ a sequence  
of elements of  the complementary set $B \setminus A$. 
When $A \subset B$, we can permute the elements of sequence $B$ so that 
it is given by $CA$,  a composite sequence of $A$ and $C$. 
Here we note that $x_m^{-}(B) =  x_m^{-}(CA)$.  
We have 
\bea
(-2) x_m^{-}(CA) \, \Omega & = & [h_{m-\ell}(C), x_{\ell}^{-}(A) ]  \, \Omega 
\non \\ 
& = & h_{m-\ell}(C) \, x_{\ell}^{-}(A) \Omega 
-  d_{m-\ell}(C) \, x_{\ell}^{-}(A) \Omega \non \\ 
& = & 0 \, . \non 
\eea
Here we have made use of (\ref{eq:d-alpha}). 
Thus, we obtain $x_m^{-}(B) \Omega = 0$. 
\end{proof}

In Appendix C, we show reduction relations for $a \ne 0$, 
which generalize (\ref{eq:rr-ell-x}).

%
\section{Derivation of irreducibility criterion}
%

For any given highest weight vector $\Omega$, 
 we have reduction relation (\ref{eq:red-all}) with respect to  
the highest weight parameters ${\hat a}_j$ for $j= 1, 2, \ldots, r$.  
It is of an $r$th order. 
However, it is not always the case that 
the following $s$th order relation holds: 
\be 
x_{s}^{-}({\bm a}) \, \Omega = 0 \, . \label{eq:red-s}
\ee
Here we recall that $s \le r$. 

We shall show that $U \Omega$ is irreducible if 
the $s$th order relation (\ref{eq:red-s}) holds. 
Here we recall that the highest weight parameters ${\hat a}_k$ are given by 
distinct parameters $a_j$ with multiplicities $m_j$ 
for $j= 1, 2, \ldots, s$ where $s \le r$ 
and ${\bm a}=(a_1, a_2, \cdots, a_s)$. Here we also note that 
condition (\ref{eq:red-s}) is similar to the criterion for 
a matrix to have no Jordan blocks.  

It is easy to show that the $s$th order relation (\ref{eq:red-s})
is necessary for $U \Omega$ to be irreducible. Let us show that 
$x_n^{+} x_s^{-}({\bm a}) \Omega = 0$  for all $n \in {\bm Z}$. 
Here, from proposition \ref{prop:hmalpha} we have  
$$
x_n^{+} x_s^{-}({\bm a}) \Omega 
= h_{n+s}({\bm a}) \Omega = 
\sum_{j=1}^{r} {\hat a}_j^{n} \prod_{i=1}^{s} ({\hat a}_j - a_i) 
\Omega =0 \, . 
$$
Therefore, if $x_s^{-}({\bm a}) \Omega \ne 0$, then 
 $U x_s^{-}({\bm a}) \Omega$ is a proper 
submodule of $U \Omega$, 
and hence $U \Omega$ is reducible.  It thus follows that 
   $x_s^{-}({\bm a}) \Omega = 0$ if $U \Omega$ is irreducible.

\begin{lem} Let ${\bm \alpha}$ and 
${\bm \beta}$ be sequences of complex parameters 
such as ${\bm \alpha}=(\alpha_1, \alpha_2, \ldots, \alpha_{\ell})$
and ${\bm \beta}=(\beta_1, \beta_2, \ldots, \beta_n)$.   
If  $x_{\ell}^{-}({\bm \alpha}) \Omega = 0$, we have  
\be 
x_{\ell+n-1}^{-}({\bm \alpha}_j {\bm \beta}) \Omega = 
\prod_{k=1}^{n} (\alpha_j - \beta_k) \, 
x_{\ell-1}^{-}({\bm \alpha}_j) \Omega \, , 
\quad \mbox{for} \, \, j=1, 2, \ldots, s.  
\label{eq:ajbeta} 
\ee
\label{lem:ajbeta} 
\end{lem} 
\begin{proof} 
We show it by induction on $n$. Here we note that $n$ denotes 
the number of parameters $\beta_k$ in eq. (\ref{eq:ajbeta}).  
For $n=1$ we note  
$$
x_{\ell}^{-}({\bm \alpha}_j (\beta_1)) - x_{\ell}^{-}({\bm \alpha}) 
= (\alpha_j - \beta_1) \, \, x_{\ell-1}^{-}({\bm \alpha}_j) \, . 
$$ 
Thus, we have $x_{\ell}^{-}({\bm \alpha}_j (\beta_1)) \Omega = 
(\alpha_j - \beta_1) \, \, x_{\ell-1}^{-}({\bm \alpha}_j) \Omega$ 
if $x_{\ell}^{-}({\bm \alpha}) \Omega = 0$. 
Let us assume (\ref{eq:ajbeta}) in the case of $n-1$. 
Here we recall ${\bm \beta}_n=(\beta_1, \ldots, \beta_{n-1})$. 
We note the following: 
\be 
x_{\ell+ n-1}^{-}({\bm \alpha}_j {\bm \beta}) - 
x_{\ell+n-1}^{-}({\bm \alpha} {\bm \beta}_n) 
= (\alpha_j - \beta_n) \, \, 
x_{\ell+n-2}^{-}({\bm \alpha}_j {\bm \beta}_n) \, .  
\label{eq:x-x=x}
\ee
It follows from lemma \ref{lem:vanishAB} 
that $x_{\ell+n-1}^{-}({\bm \alpha} {\bm \beta}_n) \Omega = 0$  
if $x_{\ell}^{-}({\bm \alpha}) \Omega = 0$. 
Thus, applying each side of eq. (\ref{eq:x-x=x})    
to $\Omega$,  we have (\ref{eq:ajbeta}) in the case of $n$. 
\end{proof}

\begin{lem} 
If $x_{s}^{-} ({\bm a}) \Omega = 0$,   
 we have the following relations for $j=1, 2, \ldots, s$: 
\be 
x_n^{+} \left( \rho^{-}_{j}({\bm a}; s) \right)^{(m_j+1)} \Omega = 0 
\, , \quad \mbox{\rm for} \quad n \in {\bm Z} \, . 
\label{eq:xn}
\ee
\end{lem}
\begin{proof} 
 Using eqs. (\ref{eq:AB}) we have 
\be  
x_0^{+} \, (\rho_j^{-}({\bm a}; s))^{(m_j+1)} \, \Omega 
= \left( \rho_{j}^{-}({\bm a}; s) \right)^{(m_j)} h_{s-1}({\bm a}_j) \, \Omega 
- \left( \rho_{j}^{-}({\bm a}; s) \right)^{(m_j-1)} 
x_{2s-2}^{-} ({\bm a}_j {\bm a}_j) \Omega \, . 
\label{eq:x0+}
\ee
In terms of $a_{kj}=a_k - a_j$, we have 
\be 
h_{s-1}({\bm a}_j) \, \Omega = m_j \, 
\left( \prod_{k=1; k \ne j}^{s} a_{jk} \right) \, \Omega \, . 
\label{eq:hs-1}
\ee 
Making use of lemma \ref{lem:ajbeta} with ${\bm \alpha}$ given by ${\bm a}$ 
we have 
\be 
x_{2s-2}^{-} ({\bm a}_j {\bm a}_j) \, \Omega 
= \left( \prod_{k=1; k \ne j}^{s} a_{jk} \right) \, 
x_{s-1}^{-}({\bm a}_j) \, \Omega 
= \left( \prod_{k=1; k \ne j}^{s} a_{jk} \right) \, 
\rho_{j}^{-}({\bm a}; s) \, \Omega \, . 
\label{eq:x2s-2}
\ee 
Putting (\ref{eq:hs-1}) and (\ref{eq:x2s-2}) into 
(\ref{eq:x0+}), we obtain eq. (\ref{eq:xn}) for $n=0$.
For any given integer $n$,      
we have      
\bea  
x_n^{+}(a_j) \, (\rho_j^{-}({\bm a}; s))^{(m_j+1)} \, \Omega 
& = & \left( \rho_{j}^{-}({\bm a}; s) \right)^{(m_j)} 
h_{n+s-1}({\bm a}_j (a_j)) \, \Omega \non \\
&  & - \left( \rho_{j}^{-}({\bm a}; s) \right)^{(m_j-1)} 
x_{n+2s-2}^{-} ({\bm a}_j {\bm a}_j (a_j)) \Omega  \non \\ 
& = & 0 \, . 
\label{eq:xn-a}
\eea
Here we note that $h_{n+s-1}({\bm a}_j (a_j)) \Omega=0$, and 
it follows from  lemma \ref{lem:vanishAB} that 
we have  $x_{n+2s-2}^{-} ({\bm a}_j {\bm a}_j (a_j)) \Omega=0 $. 
We now derive from eq. (\ref{eq:xn-a})
the following recursive relation with respect to $n$: 
\be 
x_n^{+} \, (\rho_j^{-}({\bm a}; s))^{(m_j+1)} \, \Omega 
= a_j x_{n-1}^{+} \, (\rho_j^{-}({\bm a}; s))^{(m_j+1)} \, \Omega \, . 
\label{eq:recursive}
\ee
Making use of relation (\ref{eq:recursive}), 
we derive (\ref{eq:xn}) for all $n$ from that of $n=0$. 
\end{proof}

\begin{lem} 
If $x_{s}^{-} ({\bm a}) \Omega = 0$,  we have 
\be 
{[} (x_0^{+})^{(n)}(x_1^{-}(a_j))^{(n)}, 
(\rho_j^{-} ({\bm a}; s))^{(m_j+1)} {]} \Omega = 0 \, , \quad  
\mbox{for} \quad  n \in {\bm Z}_{>0} \, .  
\label{eq:commute}
\ee
\label{lem:commute}
\end{lem}
\begin{proof}
We show it by induction on $n$. 
First, we have for any positive integer $k$ 
\be 
{[} h_k((a_j)^k), \left( \rho_j^{-}({\bm a}; s) 
\right)^{(m_j+1)} {]} \, \Omega = 
(-2) \left( \rho_j^{-}({\bm a}; s) \right)^{(m_j)} 
x_{s+k-1}^{-} ({\bm a} (a_j)^{k-1}) \, \Omega = 0 \, .  
\label{eq:first}
\ee
Secondly, we derive relation (\ref{eq:commute}) 
in the case of $n=1$ as follows: 
\bea 
& & {[} (x_0^{+})(x_1^{-}(a_j)), 
(\rho_j^{-} ({\bm a}; s))^{(m_j+1)} {]} \, \Omega 
\non \\
& = &  {[} h_1(a_j), \left( \rho_j^{-} ({\bm a}; s) \right)^{(m_j+1)} {]} 
\, \Omega 
+ x_1^{-}(a_j) x_0^{+} \, \left( \rho_j^{-} ({\bm a}; s) \right)^{(m_j+1)} \, 
\Omega 
\non \\
& = & 0 \, . \non 
\eea
Thirdly, assuming relations (\ref{eq:commute}) for 
the cases of $n < p$, we show the case of $n=p$ as follows:  
Using $({\rm B}_p)$ of lemma \ref{lem:ABC} with $a=a_j$ and $\ell=0$, 
for some element $x^{+}$ of  $U({\cal B}_{0}) {\cal B}_{0}^{+}$  
we have 
\be   
({ x}_{0}^{+})^{(p)} ({  x}_{1}^{-}(a_j))^{(p)}  
 = (-1)^{p-1} {\frac 1 p} \, h_{p}((a_j)^p) + 
  {\frac 1 p} \, \sum_{k=1}^{p-1} (-1)^{k-1} { h}_{k}((a_j)^k) 
 ({  x}_{0}^{+})^{(p-k)} 
 ({  x}_{1}^{-}(a_j))^{(p-k)}  
 + x^{+} \, .  
\label{eq:p}
\ee
Substituting (\ref{eq:p}) into the commutator, we have 
\bea 
& & {[} (x_0^{+})^{(p)}(x_1^{-}(a_j))^{(p)}, 
(\rho_j^{-} ({\bm a}; s))^{(m_j+1)} {]} \Omega 
\non \\
& & = 
(-1)^{p-1} {\frac 1 p} \, {[} h_{p}((a_j)^p), 
 (\rho_j^{-} ({\bm a}; s))^{(m_j+1)} {]} \Omega  + 
 {[} x^{+}, (\rho_j^{-} ({\bm a}; s))^{(m_j+1)} {]} \Omega \non \\
& & \quad +  
  {\frac 1 p} \, \sum_{k=1}^{p-1} (-1)^{k-1} \, 
 {[} { h}_{k}((a_j)^k) \, ({  x}_{0}^{+})^{(p-k)} 
 ({  x}_{1}^{-}(a_j))^{(p-k)},   
(\rho_j^{-} ({\bm a}; s))^{(m_j+1)} {]} \Omega \non \\
&  & = 0 \, . 
\label{eq:commute-p}
\eea
Here we have used (\ref{eq:first}) and (\ref{eq:xn}). 
We thus obtain (\ref{eq:commute}). 
\end{proof}

\begin{prop}  
If $x_{s}^{-} ({\bm a}) \Omega = 0$,  we have 
\be 
\left(\rho_j^{-}({\bm a}; s)\right)^{m_j+1} \, \Omega= 0 \, . 
\label{eq:vanish} 
\ee
\label{prop:m+1}
\end{prop}
\begin{proof}
We recall $h_0 \Omega = 
r \Omega$. We have  
\be 
\left( x_1^{-}(a_j) \right)^{r-m_j} \, 
\left( \rho_{j}^{-}({\bm a}; s) \right)^{m_j+1} \Omega = 0  
\label{eq:r-mj}
\ee
Here we note $(r-m_j)+ (m_j + 1) > r$, and there is no 
nonzero element in the sector of $h_0=-r-2$ in $U \Omega$. 
 We now apply $(x_0^{+})^{(r-m_j)}(x_1^{-}(a_j))^{(r-m_j)}$ to 
$(\rho_j^{-}({\bm a}; s))^{(m_j+1)} \, \Omega$. 
The product therefore vanishes: 
\be  
(x_0^{+})^{(r-m_j)} (x_1^{-}(a_j))^{(r-m_j)} \, \times \, 
(\rho_j^{-}({\bm a}; s))^{(m_j+1)} \, \Omega= 0 \, . 
\label{eq:product}
\ee 
It follows from  commutation relation (\ref{eq:commute})   
that the left-hand-side of (\ref{eq:product}) is given by 
\bea 
& & (\rho_j^{-}({\bm a}; s))^{(m_j+1)} \,  \times \, 
(x_0^{+})^{(r-m_j)}(x_1^{-}(a_j))^{(r-m_j)} \, \Omega \non \\
& = & 
\sum_{1 \le k_1 < \cdots < k_{n} \le r} 
({\hat a}_{k_1} - a_j) \cdots ({\hat a}_{k_n} - a_j) \,  \, \times \,  \, 
(\rho_j^{-}({\bm a}; s))^{(m_j+1)} \, \Omega \non \\
& = & \left( \prod_{k=1; k \ne j}^{s} 
a_{kj}^{m_k} \right) \, \times \,  \, 
(\rho_j^{-}({\bm a}; s))^{(m_j+1)} \, \Omega 
 \, ,  \label{eq:prod}
\eea
where $n$ denotes $r-m_j$ (see also lemma \ref{lem:xx-lambda-a}). Here, 
the number of such parameters 
${\hat a}_k$ that are not equal to $a_j$ is given by $n=r-m_j$, 
and hence we have the last line of (\ref{eq:prod}). In fact,    
we have only one set of integers $k_1 < \cdots < k_n$ 
that make the following product nonzero:  
$({\hat a}_{k_1} - a_j) \cdots ({\hat a}_{k_n} - a_j)$ for $n=r-m_j$. 
Since the product is nonzero,  i.e. 
$\prod_{k=1; k \ne j}^{s} a_{kj}^{m_k} \ne 0$, 
we obtain (\ref{eq:vanish}). 
\end{proof} 

Let us define the binomial coefficients for 
integers $n$ and $k$ with $n \ge k \ge 0$ as follows:  
\be 
\left( 
\begin{array}{c} 
n \\
k
\end{array}
\right)
= {\frac {n!} {(n-k)! k !}} \,  .  \label{eq:binomial}
\ee
\begin{lem} 
Suppose that we have $x_{s}^{-}({\bm a}) \Omega = 0$.   
Let $n$ be a nonnegative integer.  
We take such nonnegative integers $\ell_j$ and $k_j$  
for $j=1, 2, \ldots, s$ that satisfy 
$\ell_1 + \cdots + \ell_s=k_1 + \cdots + k_s=n$. Then, we have 
\be 
\prod_{j=1}^{s} 
\left( \rho^{+}_{j}({\bm a}; s) \right)^{(\ell_j)}
\, \times \, 
\prod_{j=1}^{s} 
\left( \rho^{-}_{j}({\bm a}; s) \right)^{(k_j)} \, \Omega 
= \prod_{j=1}^{s} 
\left( \delta_{\ell_j, k_j} 
\left( 
\begin{array}{c}  
m_j \\ 
k_j 
\end{array} 
\right) 
 \prod_{t=1; t \ne j}^{s}  
a_{j t}^{2 k_j} 
\right) 
\, \Omega 
\, . 
\label{eq:inner-product} 
\ee
Here $\delta_{j,k}$ denotes the Kronecker delta. 
\label{lem:inner-product}
\end{lem}
\begin{proof} 
First, it is easy to show the following:     
\be  
\rho^{+}_{i}({\bm a}; s) 
\, \times \,  
\left( \rho^{-}_{j}({\bm a}; s) \right)^{(k_j)} \, \Omega 
= 
\delta_{i, j} \,
(m_j - k_j +1) \,  
\prod_{t=1; t \ne j}^{s}  
a_{j t}^{2}  \,  
 \left( \rho^{-}_{j}({\bm a}; s) \right)^{(k_j-1)} 
\, \Omega 
\, . 
\label{eq:rho+rho-} 
\ee
Relation (\ref{eq:rho+rho-}) is derived through a similar method  
such as the case of eqs. (\ref{eq:xn}). 
By induction on $k_j$ and making use of (\ref{eq:rho+rho-}), 
 we can show the following: 
\be 
\left( \rho^{+}_{j}({\bm a}; s) \right)^{(k_j)}
\, \times \, 
\left( \rho^{-}_{j}({\bm a}; s) \right)^{(k_j)} \, \Omega 
=  
\left( 
\left( 
\begin{array}{c}  
m_j \\ 
k_j 
\end{array} 
\right) 
 \prod_{t=1; t \ne j}^{s}  
a_{j t}^{2 k_j} 
\right) 
\, \Omega 
\, . 
\label{eq:inner} 
\ee
We thus obtain (\ref{eq:inner-product}) . 
\end{proof}

\begin{prop} 
If $x_s^{-}({\bm a}) \Omega = 0$,   the set of vectors 
$\prod_{j=1}^{s} \left( \rho_{j}^{-}({\bm a}; s) \right)^{(k_j)}\, \Omega$ 
where $0 \le k_j \le m_j$ for 
$j= 1, 2, \ldots, s$, gives a basis of $U \Omega$.    
\label{prop:basis}
\end{prop}
\begin{proof}
It is derived from lemma \ref{lem:inner-product} that 
vectors 
$\prod_{j=1}^{s} \left( \rho_{j}^{-}({\bm a}; s) \right)^{(k_j)}\, \Omega$ 
are nonzero,  if we have $0 \le k_j \le m_j$ for $j= 1, 2, \ldots, s$.   
It follows from proposition \ref{prop:m+1}, 
the definition of $\rho_j^{-}({\bm a}; s)$ 
and the condition: $x_{s}^{-}({\bm a}) \Omega = 0$ 
that every vector $v_n$ in the sector of  $h_0=r - 2n$ 
is expressed as a linear combination of vectors  
$\left( \rho_{1}^{-}({\bm a}; s) \right)^{(k_1)} 
 \cdots \left( \rho_{s}^{-}({\bm a}; s) \right)^{(k_s)} \,  \Omega$  
over sets of such integers $k_1, \ldots, k_s \in {\bm Z}_{\ge 0}$ 
that satisfy  $k_1+ \cdots +k_s = n$ 
where $0 \le k_j \le m_j$ for 
$j= 1, 2, \ldots, s$.  We thus have 
\be 
v_n = \sum_{k_1=0}^{m_1} \sum_{k_2=0}^{m_2} \cdots \sum_{k_s=0}^{m_s} 
\, \delta_{k_1+ \cdots + k_s, n} \, 
C_{k_1, \cdots, k_s} \, \prod_{j=1}^{s} 
\left( \rho^{-}_{j}({\bm a}; s) \right)^{(k_j)} \, \Omega \, . 
\label{eq:vnC} 
\ee
In fact, 
we have from lemma \ref{lem:inner-product} 
$$ 
\prod_{j=1}^{s} (\rho_{j}^{+} ({\bm a}; s))^{(\ell_j)} \, \times \, v_n 
= C_{\ell_1, \cdots, \ell_s} \, \times \, 
\prod_{j=1}^{s} 
\left(
\left( 
\begin{array}{c}  
m_j \\ 
\ell_j 
\end{array} 
\right) 
 \prod_{t=1; t \ne j}^{s}  
a_{jt}^{2 \ell_j} 
\right) \, 
\, \Omega 
$$
It thus follows that if $v_n=0$, 
all the coefficients $C_{k_1, \cdots, k_s}$ are zero. 
\end{proof}

\begin{cor} 
If $x_s^{-}({\bm a}) \Omega = 0$,   the dimension of 
$U \Omega$ is given by 
\be
{\rm dim} U \Omega = \prod_{j=1}^{s} (m_j+1) \, . 
\ee
\label{cor:dimension}
\end{cor}

\begin{th1} 
Let $\Omega$ be a highest weight vector with 
highest weight parameters ${\hat a}_k$ for $k= 1, 2, \ldots, r$. 
Here, ${\hat a}_k$ 
are given by distinct parameters $a_j$ with multiplicities 
$m_j$ for $j=1, 2, \ldots, s$, which are expressed as   
${\bm a}=(a_1, a_2, \ldots, a_s)$. We assume that 
 $\Omega$ generates a finite-dimensional representation, 
 and we denote it by $U \Omega$.  
Then,  $U \Omega$  is irreducible if and only if 
$x_{s}^{-}({\bm a}) \Omega = 0$.   
\label{th:criterion}
\end{th1}
\begin{proof} 
We now show that if $x_{s}^{-}({\bm a}) \Omega = 0$, 
every nonzero vector of $U \Omega$ has such an element 
of the loop algebra 
that maps it to $\Omega$. Suppose that there is a nonzero vector $v_n$ 
in the sector $h_0= r-2n$ of $U \Omega$ that has no such element. 
Then, we have  
 \be \left( \sum_{k_1, \ldots, k_n} 
C_{k_1, \ldots, k_n}  x_{k_1}^{+}\cdots x_{k_n}^{+}  \right) 
\, \, v_n = 0 
\ee 
for all linear combinations of 
monomial elements $x_{k_1}^{+}\cdots x_{k_n}^{+}$.  
Let us express $v_n$ in terms of basis vectors 
$\left( \rho_{1}^{-}({\bm a}; s) \right)^{(k_1)} 
\cdots \left( \rho_{s}^{-}({\bm a}; s) \right)^{(k_s)} \, \Omega$ 
with coefficients $C_{k_1, \ldots, k_s}$ 
where $k_1, \ldots, k_s \in {\bm Z}_{\ge 0}$ 
satisfy $k_1 + \cdots + k_s = n$, as shown in (\ref{eq:vnC}).  
We multiply $v_n$ with $\left( \rho_{1}^{+}({\bm a}; s) \right)^{(j_1)} 
\cdots \left( \rho_{s}^{+}({\bm a}; s) \right)^{(j_s)}$ for 
a set of non-negative integers $j_1, \ldots, j_s$ 
satisfying $j_1 + \cdots + j_s =n$.  
Then, it follows from eq. (\ref{eq:inner-product}) 
that the coefficient $C_{j_1, \ldots, j_s}$ vanishes. 
We thus have shown that all the coefficients $C_{j_1, \ldots, j_s}$ vanish. 
However, this contradicts with the assumption that 
$v_n$ is nonzero. It therefore follows that $v_n$ has such an element 
that maps it to $\Omega$. 
\end{proof}

%
\section{Reducible highest weight representations}
%

Let us recall that $\Omega$ denotes a non-zero 
highest weight vector with highest 
weight $d_k$ where  $d_0=r$ 
and $U \Omega$ is finite dimensional.   
In \S 6 we shall formulate a method for constructing 
practically all 
finite-dimensional representations 
generated by such highest weight vectors 
that have the same  highest weight parameters 
 ${\hat {\bm a}}=({\hat a}_1, \ldots, {\hat a}_r)$.  
Throughout \S 6, we assume that the  
highest weight parameters ${\hat a}_1, \ldots, {\hat a}_r$ 
are given by distinct parameters $a_j$ with multiplicities $m_j$ 
for $j=1, 2, \ldots, s$. 
We denote them by ${\bm a} =(a_1, a_2, \ldots, a_s)$.
Here we also recall rule (\ref{eq:hat-a}). 

\subsection{Summary of the irreducible case}

As shown in (\ref{eq:red-all}), we always have   
$$  
x_{r}^{-} ({\hat {\bm a}}) \Omega = 0 \, . 
$$
However, if we have the following relation:  
$$ 
x_{s}^{-}({\bm a}) \, \Omega = 0 \, , 
$$ 
then it follows from theorem \ref{th:criterion} that 
the highest weight representation $U \Omega$ of $\Omega$ is irreducible, 
and the dimensionality of $U \Omega$ is given by 
\be 
{\rm dim} \, U \Omega = \prod_{j=1}^{s} (m_j+1) \, . 
\ee
Here we recall the symbol: ${\bm a} = (a_1, a_2, \ldots, a_s)$.

If $x_{s}^{-}({\bm a}) \, \Omega \ne 0$, then  
$U \Omega$ is reducible.  In order to 
determine the dimensions of reducible highest weight representations in 
$U \Omega$, we shall discuss   
further conditions of $\Omega$ in the next subsection.

\subsection{Highest weight representations with the same given highest weight}
\subsubsection{Some notation}

Let  $i_1, \ldots, i_m$ be a set of integers satisfying 
$1 \le i_1 < \ldots < i_m \le r$. 
We consider a subsequence of ${\hat {\bm a}}$ 
with respect to  $i_1, \ldots, i_m$ and denote it by $A$, i.e. 
$A=({\hat a}_{i_1}, \ldots, {\hat a}_{i_m})$.  
We denote by ${\hat {\bm a}}\setminus A$ 
such a subsequence of ${\hat {\bm a}}$ 
that is obtained by removing ${\hat a}_{i_1}, \ldots, {\hat a}_{i_m}$ 
from sequence ${\hat {\bm a}}$: 
\be 
{\hat {\bm a}}\setminus A = 
({\hat a}_1, \ldots, {\hat a}_{i_1-1}, {\hat a}_{i_1+1}, \ldots, 
{\hat a}_{i_2-1}, {\hat a}_{i_2+1}, \ldots, {\hat a}_r) \, .   
\ee 
We also express it as ${\hat {\bm a}}_A$, briefly. 
We now define operators $w_A({\hat {\bm a}})$ by  
\be 
w_{A}({\hat {\bm a}})= x_{r-m}^{-}({\hat {\bm a}}_A) \, . 
\ee
Le us consider subsequence $A=((a_1)^{k_1}(a_2)^{k_2} \cdots (a_s)^{k_s})$, 
in which there are $k_j$ copies of parameters $a_j$ for 
$j=1, 2, \ldots, s$. We denote it simply as $1^{k_1} 2^{k_2} \cdots s^{k_s}$. 
If $k_i \ne 0$ only for $i=j$ and $k_j=k$, 
we express it as $j^{k}$, i.e. we have 
\be 
j^{k} = (a_j)^{k} \, . 
\ee

Let us now consider the following operators: 
$\rho^{-}_{j}({\bm a}; s)$ for $j=1, 2, \ldots, s$,  and  
$w_{j^{k}}({\hat {\bm a}})$ for $k =1, 2, \ldots, m_j-1$ and
 $j=1, 2, \ldots, s$.  Here we note that there are 
$r$ operators in total, since we have 
$s+ (m_1-1) + \cdots + (m_s-1) = m_1 + \cdots + m_s =r$. 
\begin{lem} 
Let  $n$ be an integer satisfying $0 \le n \le r$.  
Every vector $v_n$ in the sector of $h_0=r - 2n$ of $U \Omega$ 
is expressed as a linear combination of monomial vectors 
consisting of products of $\rho_j^{-}({\bm a}; s)$ 
and $w_{j^k}({\hat {\bm a}})$ acting on $\Omega$ 
for $j=1, 2, \ldots, s$,  and  
 $k =1, 2, \ldots, m_j-1$. 
 \label{lem:sector-reducible}
\end{lem} 
\begin{proof} 
By the Poincar{\' e}-Birkhoff-Witt theorem \cite{Jacobsen} 
applied to $U(L(sl_2))$,  
every vector $v_n$  in the subspace of weight $r - 2n$ of $U \Omega$ is 
expressed as a linear combination of monomial vectors   
$ \prod_{t=1}^{n}{x}_{j_t}^{-} \, \, \Omega$ 
for some sets of integers $j_1 \le \cdots \le j_n$. 
It follows from the $r$th order reduction relations 
(\ref{eq:red-all}) (i.e. (\ref{eq:rr-ell-x}))
that every factor $x_{j_t}^{-}$ of the monomial vectors 
${x}_{j_1}^{-} \cdots {x}_{j_n}^{-} \, \Omega$ 
 is expressed as a linear combination 
of $\rho_j^{-}({\bm a}; s)$ and $w_{j^k}({\hat {\bm a}})$. 
We therefore obtain lemma \ref{lem:sector-reducible}. 
\end{proof}

Let $\Sigma$ be a sequence of such subsequences 
of ${\hat {\bm a}}$ that are of the form of $(a_j)^{k}$ where integer $k$ 
satisfies $1 \le k \le m_j-1$.  For notational convenience, 
we also regard $\Sigma$ as a set. 
If a sequence $A$ is a component of 
$\Sigma$, we express it as follows: $A \in \Sigma$.  
We now take the product of $w_A({\hat {\bm a}})$ over $A \in \Sigma$, 
and apply it to  $\Omega$. 
We denote it by $\omega_{\Sigma}$ as follows:  
\be 
\omega_{\Sigma} = \left( \prod_{A \in \Sigma} w_A({\hat {\bm a}}) 
\right) \, \Omega \, . 
\ee
Let $\ell_j$ be a non-negative integers for $j=1, 2, \ldots, s$.  
We take  sequences of $\ell_j$ integers 
${\bm k}(\ell_j)=(k_j(1), k_j(2), \ldots, k_j(\ell_j))$ which satisfy  
$1 \le k_j(1) \le \cdots \le k_j(\ell_j) < m_j$ 
for $j=1, 2, \ldots, s$. Here we assume that if $\ell_j=0$, 
${\bm k}_j(\ell_j)$ is given by an empty set. 
We define $\Sigma({\bm k}_j(\ell_j))$  by    
\be 
\Sigma({\bm k}_j(\ell_j))  
= \left(j^{k_j(1)}, j^{k_j(2)}, \ldots, j^{k_j(\ell_j)} \right) \, . 
\ee 
We introduce the following symbol:   
$\omega_{\Sigma({\bm k}_j(\ell_j))} = w_{j^{k_j(1)}}({\bm {\hat a}}) 
\cdots w_{j^{k_j(\ell_j)}}({\bm {\hat a}}) \, \Omega$. 
We denote by ${\bm k}({\bm \ell})$ a set of sequences as follows: 
\be 
{\bm k}({\bm \ell})  
= ({\bm k}_1(\ell_1), {\bm k}_2(\ell_2), \ldots, {\bm k}_s(\ell_s) ) \, . 
\ee 
Here ${\bm \ell}=(\ell_1, \ell_2, \ldots, \ell_s)$. 
Furthermore, we denote by $\Sigma({\bm k}({\bm \ell}))$ 
the following sequence of $\Sigma({\bm k}_j(\ell_j))$s:     
\be 
\Sigma({\bm k}({\bm \ell})) = 
\bigg( \Sigma({\bm k}_1(\ell_1)),  \cdots, \Sigma({\bm k}_s(\ell_s)) \bigg). 
\ee
We now define $\omega_{\Sigma({\bm k}({\bm \ell}))}$ by 
the following vector: 
\be 
\omega_{\Sigma({\bm k}({\bm \ell}))} 
=\prod_{j=1}^{s} \left( w_{j^{k_j(1)}}({\bm {\hat a}}) 
\cdots w_{j^{k_j(\ell_j)}}({\bm {\hat a}}) \right) \, \Omega . 
\ee

For some examples,  
we shall show that the highest weight vectors of irreducible 
quotients of submodules in $U \Omega$ are given by 
vectors $\omega_{\Sigma({\bm k}({\bm \ell}))}$. 
Here we have the following.    
\begin{conj}
Every irreducible quotient of submodules in $U \Omega$ has  
a highest weight vector of the form $\omega_{\Sigma({\bm k}({\bm \ell}))}$.  
\end{conj}

\subsubsection{Useful lemmas and propositions}

\begin{df} 
Let $V$ be a submodule of $U \Omega$. 
We say that $\omega \in U \Omega$ is the highest weight modulo $V$ if 
$V \subset U \omega$ and we have  
the following conditions: 
\bea 
x_n^{+} \, \omega & = & 0 \quad {\rm mod} \, \, V  
\quad \mbox{for} \quad n \in {\bm Z} \, , \label{eq:ann-mod} \\ 
h_n \,  \omega & = & {\tilde d}_n \omega \quad 
 {\rm mod} \, \,  V  
\quad \mbox{for} \quad n \in {\bm Z} \, . \label{eq:diag-mod} 
\eea
Here eigenvalues ${\tilde d}_n$ are given by some complex numbers. 
\end{df}
Here we remark that   
if $\omega$ is the highest weight modulo $V$ for a submodule $V$ of $U \Omega$ 
and we denote $U \omega$ by $W$, 
then the dimension of $W$ is given by   
$$ 
{\rm dim} W = {\rm dim} W/V + {\rm dim} V  
$$

Let us extend the definition of the binomial coefficient (\ref{eq:binomial})
into the case of negative integers $n$ as follows:  
\be 
\left( 
\begin{array}{c} 
n \\
k
\end{array}
\right)
= {\frac {n (n-1) \cdots (n-k+1)} {k !}} 
\quad \mbox{\rm for} \, \, n \in {\bm Z} \, .  
\ee
We also denote it by $_nC_k$. 
Applying $h_n$ to $w_{j^{k}}({\hat {\bm a}})\Omega$,  we have 
linear combinations of vectors  $w_{j^{k-t}}({\hat {\bm a}})\Omega$ for
 $t=1, 2, \ldots, k-1$. 
%
%
\begin{lem} 
For integers $k$ and $k^{'}$ satisfying 
$1 \le k, k^{'} \le m_j-1$, we have the following:  
\bea 
x_n^{+} \, w_{j^{k}}({\hat {\bm a}})\Omega & = & 0 \quad 
\mbox{for} \, \,  n \in {\bm Z} \, ,  \label{eq:x-vanish} \\ 
x_n^{+} \, w_{j^{k}}({\hat {\bm a}}) 
w_{j^{k^{'}}}({\hat {\bm a}}) \Omega & = & 0  
\quad 
\mbox{for} \, \,  n \in {\bm Z} \quad 
\mbox{if} \quad k+k^{'} \le m_j 
\, ,  \label{eq:x-vanish-kk} \\ 
{[} h_n , w_{j^{k}}({\hat {\bm a}}) {]} \, \Omega & = & 
 (-2) \sum_{t=0; t < k}^{n} 
\left( 
\begin{array}{c} 
n \\
t
\end{array} \right) 
 \, a_j^{n-t} 
w_{j^{k-t}}({\hat {\bm a}}) \Omega \, , \non \\ 
{[} h_{-n}, w_{j^{k}}({\hat {\bm a}}) {]} \, \Omega 
& = & 
(-2) \sum_{t=0; t < k} 
\left( 
\begin{array}{c} 
-n \\
t
\end{array} \right) 
 \, a_j^{-n-t} 
w_{j^{k-t}}({\hat {\bm a}}) \Omega \, , 
\non \\ 
& & \qquad \mbox{for} \quad n \in {\bm Z}_{> 0} \, .   
\label{eq:sum} 
\eea
\label{lem:submodule}
\end{lem}
\begin{proof} 
It is straightforward to show (\ref{eq:x-vanish}). 
Let us now show (\ref{eq:x-vanish-kk}). If $m_j \ge k+k^{'}$, we have    
\bea 
x_n^{+} \, w_{j^{k}}({\hat {\bm a}}) 
w_{j^{k^{'}}}({\hat {\bm a}}) \Omega & = & 
h_{r-k +n}({\hat {\bm a}} \setminus j^{k}) 
w_{j^{k^{'}}}({\hat {\bm a}}) \Omega  \non \\ 
& = & 
(-2) x^{-}_{2r-k-k^{'}+n}({\hat {\bm a}}({\hat {\bm a}} \setminus j^{m_1})
\, j^{m_1-k-k^{'}}) \Omega = 0 \, . \non  
\eea 
We can show (\ref{eq:sum}) by induction on $n$. We shall discuss 
it in Appendix D. 
\end{proof}

As a corollary of lemma \ref{lem:submodule} we have the following.  
\begin{cor}
Vectors  $w_{j^{k}}({\hat {\bm a}})\Omega$ are the highest weight 
 modulo $U w_{j^{k-1}}({\hat {\bm a}})\Omega$ 
($1 \le k \le m_j-1$) for $j= 1, 2, \ldots, s$: 
\bea 
x_n^{+} \, w_{j^{k}}({\hat {\bm a}}) \Omega & = & 0 \, 
\quad \mbox{for} \, \,  n \in {\bm Z} \, , \non \\ 
h_n \,  w_{j^{k}}({\hat {\bm a}})\Omega & = & (d_n -2 a_j^n)
w_{j^{k}}({\hat {\bm a}})\Omega \quad 
 {\rm mod} \, \, U w_{j^{k-1}}({\hat {\bm a}})\Omega 
\quad \mbox{for} \, \,  n \in {\bm Z} \, . \non 
\eea
Let ${\bm k}(\ell) = (k(1), k(2), \ldots k(\ell))$ be  
a sequence of integers satisfying 
$1 \le k(1) \le k(2) \le  \cdots \le k(\ell) \le m_j-1$.  
If $k(i) + k(t) \le m_j$ for all pairs of integers $i$ and $t$ 
satisfying $1 \le i < t \le \ell$,  we have 
\be 
x_n^{+} \, \prod_{i=1}^{\ell} w_{j^{k(i)}}({\hat {\bm a}}) \, \Omega 
= 0 \,
\quad \mbox{for} \, \,  n \in {\bm Z} \, . 
\label{eq:vanish-xxx} \\ 
\ee  
\label{cor:submodule}
\end{cor} 

From lemma \ref{lem:submodule} and corollary \ref{cor:submodule}
we have the following lemma.  
\begin{lem} 
For a set of subsequences 
$\Sigma=\{ j_1^{k_1}, \ldots, j_{p}^{k_{p}} \}$ 
where $1 \le j_1 \le \cdots \le j_p \le s$ and 
$1 \le k_t \le m_{j_t}-1$ for $t=1, 2, \ldots, p$, 
we denote by $\iota_{\Sigma}(j)$ the number of such integers 
in the set $\{ j_1, j_2, \ldots, j_{p} \} $ that are equal to $j$. 
Let us denote by 
$\omega^{'}$ the following vector: 
\be 
\omega^{'} = w_{j_1^{k_1}}({\bm {\hat a}}) \cdots 
w_{j_{p}^{k_{p}}}({\bm {\hat a}}) \Omega \, .   
\ee  
If $\omega^{'}$ is the highest weight modulo $V$ where 
$V$ is a submodule of $U \Omega$,   
then the highest weight parameters ${\hat a}^{'}_j$ of $\omega^{'}$ 
are given by parameters $a_j$ with multiplicities $m_j^{'}$ 
where we have 
\be 
m_j^{'}= m_j - 2 \iota_{\Sigma}(j) \, , \quad  
\mbox{for} \quad  j=1, 2, \ldots, s, 
\ee
and eigenvalues $d_n^{'}$ of $h_n$ are given by 
\be 
d_n^{'} = d_n - 2 \sum_{i=1}^{p} a_i^n \, ,  
\ee
where $d_n^{'}$ have been defined by the following:  
\be 
h_n \omega^{'} = d_n^{'} \omega^{'} \quad  {\rm mod} \, V \, .   
\ee
Here we recall $h_n \Omega = d_n \Omega$. 
\label{lem:reduction}
\end{lem}

\begin{lem} 
Let $V$ be a submodule of $U \Omega$. 
If  vector $\omega^{'}$ is the highest weight modulo $V$ 
and it has highest weight parameters ${\hat a}_j^{'}$ 
which are given by distinct parameters 
${a}_j$ with multiplicities ${m}_j^{'}$, then  we have  
\be 
{x}_{r^{'}+1-\ell}^{-} \, \omega^{'} = \sum_{j=1}^{r^{'}} (-1)^{r^{'}-j} 
{ \lambda}_{r^{'}+1-j}^{'} \, {x}^{-}_{j-\ell} \, \omega^{'} 
\quad {\rm mod} \, V 
 \quad {\rm for} \, \, 
\ell \in {\bf Z} \, .  
\label{eq:red-rel-omega}
\ee
\label{lem:red-rel-omega}
Here ${r}^{'}$ is given by $h_0 \omega^{'} = {r}^{'} \omega^{'}$ 
and  ${\lambda}_n^{'}$ are defined by  
\be 
{\lambda}_n^{'} 
= \sum_{0 \le k_1 < \cdots < k_n \le {r}^{'}} 
{\hat a}_{k_1}^{'} {\hat a}_{k_2}^{'} \cdots {\hat a}_{k_n}^{'} \, . 
\ee
\label{lem:red-rel-omega}
\end{lem}

Similarly as (\ref{eq:xn}), 
reduction relation (\ref{eq:red-rel-omega}) 
leads to the following.    
\begin{lem} 
Let $V$ be a submodule of $U \Omega$. 
If vector $\omega^{'}$ is the highest weight modulo $V$ 
and it has such highest weight parameters that are given by 
distinct parameters ${ a}_j$ with multiplicities ${ m}_j^{'}$ for 
$j=1, 2, \ldots, s$, and furthermore 
if $x_s^{-}({\bm a}) \omega^{'}=0$ mod $V$, then we have  
\be 
\left( \rho_j^{-}({\bm a}; s) \right)^{(m^{'}_j+1)} \omega^{'} = 0 \, \, \, 
{\rm mod} \, V. 
\ee
\label{lem:power-rho}
\end{lem}

For given submodules of $U \Omega$, 
 $V_1, V_2, \ldots, V_p$,    
we denote by $V_1 + \cdots + V_p$ 
or $\sum_{j=1}^{p}V_j$ the module generated 
by $V_1 \cup V_2 \cup \cdots \cup V_p$. 
Suppose that $U \omega_{\Sigma}$ is the highest weight modulo  
submodule $\sum_{j=1}^{p} U \omega_{\Sigma_j}$ for some $\Sigma_j$ s 
 and also that the quotient $U \omega_{\Sigma}/ \sum_j U \omega_{\Sigma_j}$ 
is irreducible. Then, we can determine  
the dimension of the quotient,   
making use of theorem \ref{th:criterion}
or lemmas  \ref{lem:reduction}, \ref{lem:red-rel-omega} 
and \ref{lem:power-rho}.  

\begin{prop}
Let $j$ be an integer satisfying $1 \le j \le s$, and 
$(k_1, k_2, \ldots, k_{\ell})$  a sequence of integers 
satisfying $1 \le k_1 \le \cdots \le k_{\ell} < m_j$.  
We assume that $\omega^{'}=w_{j^{k_1}}({\bm {\hat a}}) 
\cdots w_{j^{k_{\ell-1}}} ({\bm {\hat a}}) \Omega$ is highest weight 
modulo $V$. 
If $k_{\ell} \le 2 \ell-2$, we have  
\be 
w_{j^{k_{\ell}}}({\bm {\hat a}}) \omega^{'} = 0 \quad \mbox{\rm  mod} 
\, \,  V \, .  
\label{eq:vanish-modV}
\ee
\label{prop:vanishing-product-w}
\end{prop} 
\begin{proof}
Vector $\omega^{'}$ has weight $r^{'}= r - 2\ell + 2$. 
 From lemma \ref{lem:red-rel-omega},  
 $\omega^{'}$ has the $r^{'}$th order reduction relation 
as follows: $w_{j^{2\ell-2}}({\bm {\hat a}}) \omega^{'} = 0$  
mod $V$. 
It thus follows from lemma \ref{lem:vanishAB} 
that $w_{j^{k_{\ell}}}({\bm {\hat a}}) \omega^{'} = 0$  
mod $V$. 
\end{proof}

\begin{prop}
Let $j$ be an integer satisfying $1 \le j \le s$.  For an integer    
 $\ell$ satisfying $1 \le \ell < (m_j+1)/2$,  
we define ${\bm k}^{(0)}_j(\ell)$ by 
$k_j^{(0)}(t)= 2t-1$ for $t=1, 2, \ldots, \ell$:  
\be 
{\bm k}_{j}^{(0)}(\ell)= (1, 3, \ldots, 2 \ell-1) \, . 
\ee
Then, $\omega_{\Sigma({\bm k}_{j}^{(0)}(\ell))}$ is highest weight, i.e. 
we have the following: 
\bea 
x_{n}^{+} \omega_{\Sigma({\bm k}_{j}^{(0)}(\ell))} & = & 0 \quad 
\mbox{for} \quad n \in {\bm Z} \, ,  
\label{eq:zero} \\ 
h_n \omega_{\Sigma({\bm k}_{j}^{(0)}(\ell))} & = & (d_n - 2\ell a_j^n) 
\omega_{\Sigma({\bm k}_{j}^{(0)}(\ell))} 
\quad 
\mbox{for} \quad n \in {\bm Z} \, . 
\label{eq:diagonal}  
\eea
Here we recall 
$\omega_{\Sigma({\bm k}_{j}^{(0)}(\ell))}  
= \prod_{k=1}^{\ell} w_{j^{2k-1}}({\bm {\hat a}}) \, \Omega$. 
\label{prop:2t-1}
\end{prop} 
\begin{proof} 
From (\ref{eq:diagonal}) and induction on $\ell$ we have 
(\ref{eq:zero}) for all $\ell$.
We now show (\ref{eq:diagonal}) by induction on $\ell$. 
For $\ell=1$, we have 
$\omega_{\Sigma({\bm k}_{j}^{(0)}(1))} = w_{j}({\bm {\hat a}}) \, \Omega$. 
 It is readily derived from lemma \ref{lem:submodule}  
that $w_{j}({\bm {\hat a}}) \Omega$ is highest weight. 
Let us assume (\ref{eq:diagonal}) for $\ell$. 
We have 
\bea h_n \, \omega_{\Sigma({\bm k}_{j}^{(0)}(\ell+1))} 
& = & h_n \, w_{j^{2\ell+1}}({\bm {\hat a}}) \, 
\omega_{\Sigma({\bm k}_{j}^{(0)}(\ell))}  \non \\ 
& = & 
w_{j^{2\ell+1}}({\bm {\hat a}}) \, \, h_n \, 
\omega_{\Sigma({\bm k}_{j}^{(0)}(\ell))} + 
[h_n, w_{j^{2\ell+1}}({\bm {\hat a}})] \, 
\omega_{\Sigma({\bm k}_{j}^{(0)}(\ell))}  \non \\ 
& = & (d_n - 2 \ell a_j^n) w_{j^{2\ell+1}}({\bm {\hat a}}) 
\omega_{\Sigma({\bm k}_{j}^{(0)}(\ell))} 
+ (-2) x_{r-2\ell-1+ n}^{-}({\bm {\hat a}}\setminus j^{2 \ell+1}) 
\, \omega_{\Sigma({\bm k}_{j}^{(0)}(\ell))} \, . \non  
\eea
Here we note that $r^{'}= r - 2 \ell$ for 
$\omega_{\Sigma({\bm k}_{j}^{(0)}(\ell))}$.  
Therefore, making use of 
lemma \ref{lem:ajbeta} we have 
$$
x_{r-2\ell-1+ n}^{-}({\bm {\hat a}}\setminus j^{2 \ell+1}) 
\, \omega_{\Sigma({\bm k}_{j}^{(0)}(\ell))} 
= a_j^n \, x_{r-2\ell-1}^{-}({\bm {\hat a}}\setminus j^{2 \ell+1}) 
\omega_{\Sigma({\bm k}_{j}^{(0)}(\ell))} \, . 
$$
Thus, we obtain (\ref{eq:diagonal}) for $\ell+1$. 
\end{proof}

Let us denote by $\ell_j^{\rm max}$ 
the largest integer $\ell_j$ satisfying $\ell_j < (m_j+1)/2$, 
for $j= 1, 2, \ldots, s$. 
We also denote $\ell_j^{\rm max}$  by $\ell_j^{(0)}$.   

\begin{prop} 
Let us define  vector $\omega^{\rm max}$ by 
\be 
\omega^{\rm max} 
= 
\prod_{j=1}^{s} \left( 
\prod_{k=1}^{\ell_j^{\rm max}} w_{j^{2k-1}}({\bm {\hat a}}) 
\right) \, \Omega \, . 
 \ee
Then, it is highest weight.   
The representation generated by $\omega^{\rm max}$ is 
irreducible and has the following dimension:  
\be 
\mbox{\rm dim} \left( U \omega^{\rm max} \right) = 
\prod_{k=1}^{s}(m_k + 1 - 2 \ell_k^{\rm max}) \, . 
\ee
\label{prop:max}
\end{prop}
\begin{proof}
It follows from proposition \ref{prop:2t-1} and 
lemma \ref{lem:submodule} that 
$\omega^{\rm max}$ is highest weight. 
Making use of lemma \ref{lem:reduction}, 
we evaluate the multiplicities of parameters $a_j$ 
describing the highest weight parameters of $\omega^{\rm max}$, 
and we have $m_j^{'} = m_j - 2 \ell_j^{\rm max}$ for each $j$.  
It follows from theorem \ref{th:criterion} and lemma 
\ref{lem:red-rel-omega} that the representation 
generated by $\omega^{\rm max}$ is irreducible. 
\end{proof} 

We remark that $\omega^{\rm max}$ is also expressed as follows:   
\be 
\omega^{\rm max} = \omega_{\Sigma({\bm k}^{(0)}({\bm \ell}^{(0)}))} \, . 
\ee
Here ${\bm k}^{(0)}({\bm \ell}^{(0)})$ denotes the set 
$\{{\bm k}_1^{(0)}(\ell_1^{(0)}), {\bm k}_2^{(0)}(\ell_2^{(0)}),
 \ldots, {\bm k}^{(0)}_s(\ell_s^{(0)}) \}$.

\subsubsection{Algorithm for constructing reducible 
highest weight representations}

Let us now formulate an algorithm for constructing 
irreducible quotients of submodules of $U \Omega$. 
We first construct a network of submodules 
$U \omega_{\Sigma({\bm k}({\bm \ell}))}$.  
The network consists of vertices and edges, 
where each of the vertices corresponds to a submodule 
$U \omega_{\Sigma({\bm k}({\bm \ell}))}$   
and each of the edges has an arrow  
which goes from a parental submodule 
to a daughter submodule.

For a given vector $\omega_{\Sigma({\bm k}({\bm \ell}))}$, which we call 
a parental vector,  
we construct a daughter vector $\omega_{\Sigma({\bm k}^{'}({\bm \ell}^{'}))}$ 
by the  procedures from (i) to (v) in the following. 
\begin{itemize}
\item (i) We select an integer $j$ satisfying $1 \le j \le s$.  
\item (ii)
If $\ell_j= \ell_j^{(0)}$ and $k_j(\ell_j)+1 < m_j$, then we set 
$k_{j}^{'}(\ell_j)=k_{j}(\ell_j)+1$ 
and define descendant 
$\omega_{\Sigma({\bm k}^{'}({\bm \ell}^{'}))}$ by    
$$
\omega_{\Sigma({\bm k}^{'}({\bm \ell}^{'}))} 
= w_{j^{k_j(1)}}({\bm {\hat a}}) \cdots 
w_{j^{k_{j}(\ell_j)+1}}({\bm {\hat a}}) 
\prod_{t=1; t \ne j}^{s} 
\left( w_{t^{k_t(1)}}({\bm {\hat a}}) 
\cdots w_{t^{k_{t}(\ell_t)}}({\bm {\hat a}}) \right) \, 
\Omega \, . 
$$ 
\item (iii)
If $\ell_j= \ell_j^{(0)}$ and $k_j(\ell_j)+1 = m_j$,  
then we set $\ell_j^{'}=\ell_j-1$ and 
we define descendant 
$\omega_{\Sigma({\bm k}^{'}({\bm \ell}^{'}))}$ by  
$$ 
\omega_{\Sigma({\bm k}^{'}({\bm \ell}^{'}))} = 
 w_{j^{k_j(1)}}({\bm {\hat a}}) \cdots w_{j^{k_{j}(\ell_j-1)}}({\bm {\hat a}}) 
\prod_{t=1; t \ne j}^{s}
\left( w_{t^{k_t(1)}}({\bm {\hat a}}) 
\cdots w_{t^{k_{t}(\ell_t)}}({\bm {\hat a}})  \right) \, 
\Omega \, . 
$$ 
\item (iv) If $\ell_j< \ell_j^{(0)}$ and $k_j(\ell_j) < m_j-1$, 
then we set $k_j^{'}(\ell_j)= k_j(\ell_j)+1$. Furthermore,   
we set $k_j^{'}(\ell_j+i)$= max $\{ k_j(\ell_j)+1, 2(\ell_j + i)-1 \}$ 
for $i=1, 2, \ldots, \ell_j^{(0)}-\ell_j$, 
and we set $\ell_j^{'}=\ell_j^{(0)}$.  
We define descendant 
$\omega_{\Sigma({\bm k}^{'}({\bm \ell}^{'}))}$ by  
\bea 
\omega_{\Sigma({\bm k}^{'}({\bm \ell}^{'}))} & = & 
 w_{j^{k_j(1)}}({\bm {\hat a}}) \cdots w_{j^{k_j(\ell_j-1)}}({\bm {\hat a}}) 
\times \non \\ 
& \times & w_{j^{k_j^{'}(\ell_j)}}({\bm {\hat a}}) 
\cdots 
w_{j^{k_{j}^{'}(\ell_j^{(0)})}}({\bm {\hat a}}) 
\prod_{t=1; t \ne j}^{s}
\left( w_{t^{k_t(1)}}({\bm {\hat a}}) 
\cdots w_{t^{k_{t}(\ell_t)}}({\bm {\hat a}})  \right) \, 
\Omega \, . \non 
\eea 
\item (v) If $\ell_j< \ell_j^{(0)}$ 
and $k_j(\ell_j) = m_j-1$, we set $\ell_j^{'}=\ell_j-1$.  
We define descendant 
$\omega_{\Sigma({\bm k}^{'}({\bm \ell}^{'}))}$ by  
$$ 
\omega_{\Sigma({\bm k}^{'}({\bm \ell}^{'}))} = 
 w_{j^{k_j(1)}}({\bm {\hat a}}) \cdots 
 w_{j^{k_{j}(\ell_j-1)}}({\bm {\hat a}}) 
\prod_{t=1; t \ne j}^{s}
\left( w_{t^{k_t(1)}}({\bm {\hat a}}) 
\cdots w_{t^{k_{t}(\ell_t)}}({\bm {\hat a}})  \right) \, 
\Omega \, . 
$$ 
\end{itemize} 
Here we have specified only such elements of ${\bm k}^{'}({\bm \ell}^{'})$ 
that are changed from parental one ${\bm k}({\bm \ell})$.

First, we put vector $\omega^{\rm max}$ 
at the starting point of the network. 
That is, 
we set $\ell_j=\ell_j^{(0)}$ and 
put ${\bm k}_j(\ell_j)= {\bm k}_j^{(0)}(\ell_j^{(0)})$ 
for $j=1, 2, \ldots, s$, 
as initial conditions. 
Then, we apply  the procedures of (i), (ii), $\ldots$, and (v) 
to the parental vector $\omega_{\Sigma({\bm k}({\bm \ell}))}$, 
then we derive daughter vectors 
$\omega_{\Sigma({\bm k}^{'}({\bm \ell}^{'}))}$ 
for each $j$ ( $1 \le j \le s$).  
Then, we choose one of ${\bm k}^{'}({\bm \ell}^{'})$,
and we set ${\bm k}({\bm \ell})={\bm k}^{'}({\bm \ell}^{'})$.  
We repeat the procedure again. Finally, we arrive at the 
end point of the network, where  
${\bm k}({\bm \ell})$ is given by an empty set, $(\emptyset)$.

Applying lemma \ref{lem:reduction} to 
the derived  network of submodules, we can calculate practically 
all the dimensions of reducible highest weight representations 
with the same given highest weight. 
Suppose that a submodule $V$ in the network 
has parental submodules $V_1, \ldots, V_p$. We take the quotient of 
$V$ with respect to the sum 
over all the parental submodules; 
we have $V/(V_1 + \cdots + V_p)$. 
Here we remark that if all the parental submodules 
are irreducible, the sum 
$V_1 + \cdots + V_p$ is given by the direct sum.

If the quotient $V/(V_1 + \cdots + V_p)$ 
does not vanish, then it is irreducible. 
Evaluating the multiplicities $m_j^{'}$ through lemma \ref{lem:reduction}, 
we derive the dimension of the irreducible quotient by 
corollary \ref{cor:dimension}.  
We then cut  the network into two parts such that  
one has the starting point, while another has the end point, respectively. 
Then, the subnetwork that have the end point corresponds to 
a reducible (or irreducible) highest weight representation. 
We obtain the dimension of the representation 
taking the sum of all the dimensions of the irreducible quotients in 
the remaining part of the network.

For an illustration, let us consider the case of $r=6$ 
with $(m_1, m_2)=(3, 3)$. 
Here we have $\ell_1^{\rm max}= \ell_2^{\rm max}= 1$.  
We put $\ell_1^{(0)}=\ell_2^{(0)}=1$ and start with 
${\bm k}^{(0)}({\bm \ell}^{(0)})= ((1^{1}),(2^{1}))$ where we have  
 ${\bm k}_1^{(0)}=(1^{1})$ and ${\bm k}_2^{(0)}=(2^{1})$.  
The highest weight vector $\omega^{\rm max}$ is given by 
$w_{1^1}w_{2^1} \Omega$, and it generates a four-dimensional   
irreducible module.  Here we recall proposition \ref{prop:max}. 
 Through the procedures of (i) to (v), 
we now derive all the daughter vectors. 
 For a given ${\bm k}({\bm \ell})$, we show all 
 ${\bm k}^{'}({\bm \ell}^{'})$ after the symbol, 
 '$\rightarrow$',  as follows:   
\bea 
& & 
((1^{1}),(2^{1})) \, \rightarrow ((1^{2}),(2^{1})), ((1^{1}),(2^{2})); \qquad 
((1^{2}),(2^{1})) \, \rightarrow ((\emptyset),(2^{1})), ((1^{2}),(2^{2})); \non \\ 
& &  
((1^{1}),(2^{2})) \, \rightarrow ((1^{2}),(2^{2})), ((1^{1}),(\emptyset));  
\qquad 
((1^{2}),(2^{2})) \, \rightarrow ((\emptyset),(2^{2})), ((1^{2}),(\emptyset)); 
\non \\ 
& & ((1^{1}),(\emptyset)) \, \rightarrow ((1^{2}),(\emptyset)); \qquad 
 ((\emptyset),(2^{1})) \, \rightarrow ((\emptyset),(2^{2})); \non \\
& & ((1^{2}),(\emptyset)) \, \rightarrow ((\emptyset),(\emptyset)); \qquad 
 ((\emptyset),(2^{2})) \, \rightarrow ((\emptyset),(\emptyset)) \, .  
\non 
\eea 
We have four-dimensional irreducible quotients:  
$$
U w_{1^2} w_{2^1} \Omega/ Uw_{1^1} w_{2^1} \Omega \, , \quad  
U w_{1^1} w_{2^2} \Omega/ Uw_{1^1} w_{2^1} \Omega \, , \quad 
U w_{1^2} w_{2^2} \Omega/ (Uw_{1^1} w_{2^2} \Omega +  
U w_{1^2} w_{2^1} \Omega) 
\, ,  
$$
eight-dimensional irreducible quotients:
\bea 
& & 
U w_{2^1} \Omega/ Uw_{1^2} w_{2^1} \Omega \, , \quad  
U w_{1^1} \Omega/ Uw_{1^1} w_{2^2} \Omega \, , \quad 
U w_{1^2} \Omega/ (Uw_{1^1}\Omega + Uw_{1^2} w_{2^2} \Omega) \, , 
\non \\ 
& & U w_{2^2} \Omega/ (Uw_{2^1} \Omega + U w_{1^2} w_{2^2} \Omega)
\, ,  \non 
\eea
and a 16-dimensional irreducible quotient: 
$$
U \Omega/ (Uw_{1^2} \Omega + U w_{2^2} \Omega) \, . 
$$
Here we recall the following: $w_{1^1}= x_{5}^{-}((a_1)^2 (a_2)^3)$, 
$w_{1^2}= x_{4}^{-}((a_1)^1 (a_2)^3)$, 
$w_{2^1}= x_{5}^{-}((a_1)^3 (a_2)^2)$, 
and so on.  In total, we have $64$ dimensions as follows: 
$$
4 + (4+4+4) + (8+8+8+8) + 16= 64 \, . 
$$
Here we note that it is given by $2^{6}$. 
Let us now construct an example of reducible highest weight module.  
For instance, suppose that $w_{1^1} \Omega=0$,  
$w_{1^2} w_{2^2} \Omega=0$ and $w_{2^2} \Omega=0$. We then take the sum 
of two irreducible quotients for 
${\bm k}({\bm n})=((\emptyset),(\emptyset))$  and 
$((1^{2}),(\emptyset))$ as follows: 
$$
U \Omega/ (Uw_{1^2} \Omega + U w_{2^2} \Omega) \, \oplus \,  
U w_{1^2} \Omega/ (Uw_{1^1}\Omega + U w_{1^2} w_{2^2} \Omega) \,  . 
$$
The reducible module has $16 + 8 = 24$ dimensions. We thus obtain 
a 24-dimensional reducible highest weight representation.

In summary, the algorithm consists of the following.  
We first derive sequences of irreducible quotients 
of highest weight submodules. Here they form a network of 
irreducible quotients.  
We terminate some sequences in the network at some points, 
and make it into two subnetworks.   
We then take the sum of the irreducible quotients in the 
subnetwork which has the end point of the network.   
We thus obtain a reducible highest weight submodule.

\subsubsection{Conjectured relations}

We remark that 
in the procedures from (ii) to (v), the quotient of a daughter  
submodule $U\omega_{\Sigma({\bm k}^{'}({\bm n}^{'}))}$ 
modulo the sum of all the parental submodules may have zero dimension.   
Up to $m_j=5$, we have confirmed that 
such vanishing cases are determined by using the following conjecture. 
\begin{conj} For $0 \le n \le m_j$ we have 
\be 
\sum_{k=1}^{n} k \, w_{j^{k+1}}({\bm {\hat a}}) 
w_{j^{n+1-k}}({\bm {\hat a}}) \Omega = 0 \, . 
\label{eq:conj}
\ee
\label{conj:rel}
\end{conj} 
Up to the case of $n=3$, we have shown relation (\ref{eq:conj}), 
applying relation (${\rm A}_m$) of lemma \ref{lem:ABC} 
with $m={r-n}$ to vector $w_{j^{n}} \Omega$.  

As an illustration, we consider the case of 
$r=6$ and $(m_1, m_2)=(5, 1)$. 
Here we have $\ell_1^{\rm max}= 2$ and $\ell_2^{\rm max}= 0$.  
We put $\ell_1^{(0)}=2$ and $\ell_2^{(0)}=0$.  Starting with 
${\bm k}^{(0)}= (1^1, 1^3)$ we have the following sequence 
of descendants: $(1^1, 1^3) \rightarrow  (1^1, 1^4) 
 \rightarrow (1^1) \rightarrow (1^2, 1^3) \rightarrow 
 (1^2, 1^4) \rightarrow (1^2) \rightarrow 
 (1^3, 1^3) \rightarrow (1^3, 1^4) \rightarrow 
 (1^3) \rightarrow (1^4, 1^4) \rightarrow 
 (1^4) \rightarrow (\emptyset)$. It follows from relations 
 (\ref{eq:conj}) that quotients corresponding to $(1^2, 1^3)$ and 
 $(1^3, 1^3)$ have zero dimension.  
Irreducible quotients for $(1^1, 1^3)$,  
 $(1^1, 1^4)$, $(1^2, 1^4)$, $(1^3, 1^4)$, and 
 $(1^4, 1^4)$ 
  have four dimensions, while   
irreducible quotients for $(1^1)$,  
 $(1^2)$, $(1^3)$, and $(1^4)$ have 
 eight dimensions. The dimension of the irreducible quotient 
corresponding to $(\emptyset)$ is given by $(5+1)(1+1)=12$. 
 In total, we have $2^6= 64$ dimensions as follows:  
$$
(4 + 4+ 4+ 4+ 4) + (8+ 8 + 8 +8 ) + 12 = 64 .  
$$

For $m_j > 5$ or 6, we may have 
some relations consisting of products of 
three or more $w_{j^{k}}({\bm {\hat a}})$'s 
which generalize relations (\ref{eq:conj}), and they should 
be useful for constructing reducible highest weight 
representations.

\subsection{Examples of reducible representations }

We have a conjecture that the algorithm of \S 6.2.3  leads to    
all the reducible highest weight 
representations with the same given highest weight.  
For some explicit examples,  
we shall now construct all reducible representations 
with a given highest weight by the algorithm.  

Hereafter, we write $\rho_j^{-}({\bm a}; s)$ 
and $w_j({\hat {\bm a}})$ simply as $\rho_j$ and $w_j$, 
respectively.

\subsubsection{The case of $r=3$}

Let us consider the case of $r=3$ with $m_1=2$ and $m_2=1$. 
The highest weight parameters of $\Omega$ are given by   
${\hat {\bm a}}=(a_1, a_1, a_2)$.  
It follows  that  the highest weight 
representation $U \Omega$ has four sectors of 
$h_0=3, 1, -1$, and 3, respectively. Here we recall some symbols. 
\be 
\rho_1 = x_1^{-}(a_2) \, , \quad 
\rho_2 = x_1^{-}(a_1) \, , \quad 
w_1 = x_2^{-}(a_1, a_2) \, . 
\ee
Here we recall that $x_2^{-}(a_1, a_2)$ denotes 
 $x_2^{-}(B)$ with $B=(a_1, a_2)$, and $w_1$ abbreviates $w_{1^{1}}$. 
It is easy to show from reduction relations (\ref{eq:red-all}),  
i.e. $x_n^{-}({\hat {\bm a}}) \Omega = 0$ $(n \in {\bm Z})$, that 
$x_n^{-} \Omega$ is expressed in terms of $\rho_1 \Omega$,  
$\rho_2 \Omega$, and $w_1 \Omega$ as follows: 
\be
x_n^{-} \Omega = {\frac {a_1^n} {a_{12}}} \, \rho_1 \Omega +  
{\frac {a_2^n} {a_{21}}} \, \rho_2 \Omega +
\left( {\frac {n a_1^{n-1}} {a_{12}}}
- {\frac {a_1^n -a_2^n} {a_{12}^2}} \right) \,  w_1 \Omega \quad  
\mbox{for} \, n \in {\bm Z}. 
\ee
It is straightforward to show  
\be 
x_n^{+} w_1 \Omega = 0 \, , \quad {\rm for} \quad n \in {\bf Z}  \, . 
\ee 
It thus follows that $U \Omega$ 
is reducible and indecomposable if $w_1 \Omega \ne 0$.   
From lemmas \ref{lem:red-rel-omega} and 
\ref{lem:power-rho}  we have  
\be \rho_1^3 \Omega = 0 \, , \quad  \rho_2^2 \Omega =0; 
\quad w_1^2 \Omega = 0. 
\ee
The following vectors do not vanish, if and only if $\Omega$ does not 
vanish (i.e. $\Omega \ne 0$):   
\be 
\rho_2 \Omega \ne 0 ; \quad  
\rho_1 \rho_2 \Omega \ne 0;  \quad 
\rho_1^2 \rho_2 \Omega \ne 0 \, .     
\ee
In fact, if $\rho_1^2 \rho_2 \Omega = 0$, applying 
$\rho^{+}_1({\bm a}; 2)$ and $\rho^{+}_2({\bm a}; 2)$ to it, 
we derive that $\Omega = 0$.  
From the viewpoint of lemma \ref{lem:red-rel-omega}, highest weight vector 
$w_1 \Omega$ has only one highest weight parameter $a_2$,  
i.e. $m_1^{'}=0$ and $m_2^{'}=1$,  and hence 
 we have the following reduction relation: 
\be 
\rho_1 w_1 \Omega = 0 \, .      
\ee
Furthermore, we have  $\rho_2 w_1 \Omega \ne 0$ if $w_1 \Omega \ne 0$.  
In the four sectors of $U \Omega$   
 the  basis vectors are given as follows: 

\be
\begin{array}{cccc} 
  \Omega , &  &  & {\rm for} \quad h_0=3 ; \non \\ 
 \rho_1 \Omega , & \rho_2 \Omega , &  w_1 \Omega , & 
\quad  {\rm for} \quad h_0=1 ; \non \\ 
  \rho_1^2 \Omega, & \rho_1 \rho_2 \Omega, & \rho_2 w_1 \Omega , & 
\quad  {\rm for} \quad h_0=-1 ; \non \\    
 \rho_1^2 \rho_2 \Omega , & & & \quad  {\rm for} \quad h_0=-3 \, .   
\end{array}
\ee
Consequently, we have the following result: 
\begin{prop} 
The highest weight representation with three highest 
weight parameters: (${\hat {\bm a}}$)=($a_1$, $a_1$,  $a_2$), 
 is reducible, indecomposable 
and of $2^3$ dimensions, if and only if $w_1 \Omega \ne 0$. 
It is irreducible and of 6 dimensions,  
if and only if $w_1 \Omega = 0$. 
\end{prop}

\subsubsection{The case of four highest weight parameters }

Let us consider the case of $r=4$ with $m_1=2$ and $m_2=2$.  
Here we note $(\hat {\bm a})=(a_1, a_1, a_2, a_2)$.   
 The highest weight representation $U \Omega$ has five sectors of 
$h_0 = 4, 2, 0, -2$, and $-4$. 
With the reduction relation $x_n^{-}({\hat {\bm a}})\Omega = 0$ 
($n \in {\bm Z}$) as given in eq. (\ref{eq:red-all}), 
we consider the following four operators: 
\be 
\rho_1 = x_1^{-}(a_2) \, , \quad 
\rho_2 = x_1^{-}(a_1) \, , \quad 
w_1 = x_3^{-}(a_1, a_2, a_2) \, , 
\quad w_2 = x_3^{-}(a_1, a_1, a_2) \, .  
\ee
 
Vectors $x_n^{-} \Omega$ 
for $n \in {\bm Z}$ are expressed in terms of $\rho_1 \Omega$,  
$\rho_2 \Omega$,  $w_1 \Omega$ and $w_2 \Omega$ as follows: 
\be
x_n^{-} \Omega = {\frac {a_1^n} {a_{12}}} \, \rho_1 \Omega +  
{\frac {a_2^n} {a_{21}}} \, \rho_2 \Omega + 
 {\frac 1 {a_{12}^2}} \left({n a_1^{n-1}} 
- {\frac {a_1^n -a_2^n} {a_{12}} } \right) \,  w_1 \Omega  
+ {\frac 1 {a_{12}^2}} \left({n a_2^{n-1}} 
- {\frac {a_1^n -a_2^n} {a_{12}} } \right) \,  w_2 \Omega \, . 
\ee
Making use of proposition \ref{prop:vanishing-product-w} 
(or directly from lemma \ref{lem:red-rel-omega}) that we have 
\be 
 w_1^2 \Omega= w_2^2 \Omega = 0 \, .      
\ee
Making use of lemma \ref{lem:red-rel-omega}, 
we have 
\be 
 w_1w_2 \Omega= a_{12}^2 \rho_1 w_1 \Omega
  = a_{12}^2 \rho_2 w_2 \Omega \, .      
\ee
It is also straightforward to show the following:   
\bea 
x_n^{+} w_1 \Omega & = & x_n^{+} w_2 \Omega = 0 \, ,
 \quad {\rm for} \quad n \in {\bm Z}  \, , \non \\ 
x_n^{+} \, w_1 w_2 \Omega & = & 0 \quad  
(x_n^{+} \, \rho_1 w_1 \Omega  =  x_n^{+} \rho_2 w_2 \Omega = 0) \, ,
 \quad {\rm for} \quad n \in {\bm Z}  \, . 
\eea 
It thus follows that $U \Omega$ is reducible 
if $w_1 \Omega \ne 0$, $w_2 \Omega \ne 0$ or $w_1 w_2 \Omega \ne 0$.   
 From lemma \ref{lem:power-rho} 
and proposition \ref{prop:vanishing-product-w} we also have the following: 
\be \rho_1^2 \Omega \ne 0 \, , \quad 
\rho_1^3 \Omega =0;  \quad 
\rho_2^2 \Omega \ne 0\, , \quad \rho_2^3 \Omega =0;   
\quad w_1^2 \Omega = 0. 
\ee

The basis vectors of $U \Omega$  are given by  
\be
\begin{array}{ccccccc} 
  \Omega , &  &  &  &  & & {\rm for} \quad h_0=4 ; \non \\ 
 \rho_1 \Omega , & \rho_2 \Omega , &  &  w_1 \Omega , &  w_2 \Omega , & 
& \quad  {\rm for} \quad h_0=2 ; \non \\ 
  \rho_1^2 \Omega, & \rho_1 \rho_2 \Omega, & \rho_2^2 \Omega , 
   & \rho_2 w_1 \Omega , & \rho_1 w_2 \Omega ,  & w_1w_2 \Omega , & 
\quad  {\rm for} \quad h_0=0 ; \non \\    
 \rho_1^2 \rho_2 \Omega , &  \rho_1 \rho_2^2 \Omega ,    & 
 & \rho_2^2 w_1 \Omega ,  & \rho_1^2 w_2 \Omega ,  & 
 & \quad  {\rm for} \quad h_0=-2 \, , \non \\  
 \rho_1^2 \rho_2^2 \Omega , & & & & & & \quad  {\rm for} \quad h_0=-4 \, .  
\end{array}
\ee

As an illustration, let us apply the algorithm of \S 6.2.3 for constructing 
reducible highest weight representations. 
Here we recall that $m_1=2$ and $m_2=2$. Here we have 
$\ell_1 < (m_1 + 1)/2 = 3/2$, and we obtain $\ell_1^{\rm max}=1$.   
Similarly, we have $\ell_2^{\rm max}=1$. Thus, we first consider 
$\omega^{\rm max} = U w_{1^{1}}w_{2^{1}} \Omega$. 
\begin{enumerate} 
\item 
$U w_1 w_2 \Omega$ has $r^{'}= 4 - 2 \times 2 =0$ and it has 1 dimension;  
\item 
$U w_1 \Omega/U w_1 w_2 \Omega $ has $r^{'} = 4 - 1 \times 2 =2$ where 
$({\hat a}_1^{'}, {\hat a}_2^{'}) = (a_1, a_1)$, and it has 3 dimensions 
since (2+1)= 3;  
\item
$U w_2 \Omega/U w_1 w_2 \Omega $ has $r^{'} = 4 - 1 \times 2 =2$ where 
$({\hat a}_1^{'}, {\hat a}_2^{'}) = (a_2, a_2)$, and it has  3 dimensions 
since (2+1)= 3; 
\item 
$U \Omega$ modulus $U w_1 \Omega$ and $U w_2 \Omega$ has 
$r=4$ where 
$({\hat a}_1, {\hat a}_2, {\hat a}_3, {\hat a}_4) = 
(a_1, a_1, a_2, a_2)$, and it has $(2+1) \times (2+1)= 9$, i.e. 9 dimensions. 
\end{enumerate}
We note that $w_1 - w_2 = a_{12} x_2^{-}(a_1, a_2)$. Thus, 
if $w_1 \Omega = 0$ and $w_2 \Omega = 0$, we have 
$x_2^{-}(a_1, a_2) \Omega = 0$. That is,  
$\Omega$ generates an irreducible representation.

In summary, for all possible dimensions 
of reducible and irreducible highest weight representations 
with highest weight parameters $(a_1, a_1, a_2, a_2)$,  
we have the following result: 
\begin{prop} 
If the highest weight representation with 
highest weight parameters ${\hat {\bm a}}= (a_1, a_1, a_2, a_2)$ 
is reducible, there are the following four cases:  
(i)  $w_1 \Omega \ne 0$, $w_2 \Omega \ne 0$, and $w_1 w_2 \Omega \ne 0$; 
(ii) $w_1 \Omega \ne 0$, $w_2 \Omega \ne 0$, and $w_1 w_2 \Omega = 0$; 
(iii) $w_1 \Omega \ne 0$, $w_2 \Omega = 0$, and $w_1 w_2 \Omega = 0$; 
(iv) $w_1 \Omega = 0$, $w_2 \Omega \ne 0$, and $w_1 w_2 \Omega = 0$.  
It has dimensions 16, 15, 12, 12, for cases  
(i), (ii), (iii) and (iv), respectively. 
It is irreducible if and only if 
$w_1 \Omega = 0$ and $w_2 \Omega = 0$. If it is irreducible, 
it has nine dimensions. 
 \end{prop}

\section*{Acknowledgment} 
This work is partially supported by 
Grant-in-Aid for Scientific Research (C) No. 17540351.

\appendix

%
%

\setcounter{section}{0}
 \setcounter{equation}{0} 
 \renewcommand{\theequation}{A.\arabic{equation}}
\section{Proof of proposition \ref{prop:1-dim}}

Applying the Poincar{\' e}-Birkhoff-Witt theorem \cite{Jacobsen} 
we can show that 
every vector $v$ in the subspace of weight $-r$ of $U \Omega$ is written 
as follows 
\be 
v = \sum_{k_1 \le  \cdots \le k_r} C_{k_1, \ldots, k_r} \, 
x_{k_1}^{-} \cdots x_{k_r}^{-} \Omega \, . 
\ee
Here the coefficients $C_{k_1, \ldots, k_r}$ are some complex numbers. 
Then, we obtain proposition \ref{prop:1-dim} from the following lemma: 
\begin{lem2} Let $n$ be a non-negative integer and $n \le r$. 
For any given set of integers $k_1, \ldots, k_n$, we have  
\be 
(x_0^{-})^{r-n} x_{k_1}^{-} \cdots x_{k_n}^{-} \Omega 
= A_{k_1, \ldots, k_n} (x_0^{-})^{r} \Omega  \, . 
\label{eq:r-n}
\ee
Here $A_{k_1, \ldots, k_n}$ is a complex number. 
\end{lem2} 
\begin{proof} 
We show it by induction on $n$.  The case of $n=0$ is trivial. 
Suppose that relations (\ref{eq:r-n}) hold 
for the cases $n-1$ and $n$. We show the case of $n+1$ as follows: 
We have from (\ref{eq:r-n}) in the case of $n$ the following: 
\bea 
x_{m}^{+} (x_0^{-})^{r+1-n} \prod_{j=1}^{n} x_{k_j}^{-}  \Omega  
& = &  x_{m}^{+} x_0^{-} \cdot 
A_{k_1, \ldots, k_n} (x_0^{-})^{r} \Omega \non \\ 
& = & A_{k_1, \ldots , k_n} x_{m}^{+}  \cdot 
 (x_0^{-})^{r+1} \Omega = 0 \, . \non 
\eea
Calculating the commutation relation:  
$[ x_{m}^{+}, (x_0^{-})^{(r+1-n)} \prod_{j=1}^{n} x_{k_j}^{-}]$,  
we have 
\bea 
& & ({x}_{0}^{-})^{(r-n-1)}  
\, {  x}_{m}^{-} \prod_{j=1}^{n} x_{k_j}^{-} \, 
 \Omega 
= 
{  d}_{m}
 ({  x}_0^{-})^{(r-n)}  
\prod_{j=1}^{n} x_{k_j}^{-} \, \Omega 
\non \\
& &   
+ (-2) \sum_{i=1}^{n} ({x}_{0}^{-})^{(r-n)} 
{  x}_{k_i + m+1}^{-} \prod_{j=1; j \ne i}^{n}  {  x}_{k_j}^{-} \Omega  
+ \sum_{i=1}^{n} { d}_{m+1+k_i} ({x}_{0}^{-})^{(r+1-n)} 
\prod_{j=1; j \ne i}^{n}  { x}_{k_j}^{-}  \Omega 
\non \\
& & \quad + (-2) \sum_{1 \le i_1 < i_2 \le n} 
({  x}_{0}^{-})^{(r+1-n)} {  x}_{ m + 1+ k_{i_1} + k_{i_2}}^{-} 
\prod_{j=1; j \ne i_1, i_2}^{n}  { x}_{k_j}^{-} \Omega \, . 
\eea
Denoting $m$ by $k_{n+1}$, 
we thus obtain relation  (\ref{eq:r-n}) for the case of $n+1$.  
\end{proof}

%
%

 \setcounter{equation}{0} 
 \renewcommand{\theequation}{B.\arabic{equation}}

\section{Recursive relations for Drinfeld generators}

We now show lemma \ref{lem:ind-ell}. 
We recall that $(X)^{(n)}$ denotes 
$(X)^{(n)}=X^n/n!$. 
\begin{lem2} The following recursive formula with respect to $n$
holds for products of operators 
$({  x}_{\ell}^{+})^{(n-1)} ({  x}_{1-\ell}^{-}(a))^{(n)}$:
\bea 
({  x}_{\ell}^{+})^{(n)} ({  x}_{1-\ell}^{-}(a))^{(n+1)} 
& = & {  x}_{1-\ell}^{-}(a) 
({  x}_{\ell}^{+})^{(n)} ({  x}_{1-\ell}^{-}(a))^{(n)} 
+ {\frac 1 2} \, {[} {  h}_{1}, ({  x}_{\ell}^{+})^{(n-1)} 
({  x}_{1-\ell}^{-}(a))^{(n)} {]} 
\non \\
& & \quad - ({  x}_{\ell}^{+})^{(n-1)} 
({  x}_{1-\ell}^{-}(a))^{(n+1)} {  x}_{\ell}^{+} \, , \quad 
 {\rm for} \,\,  \ell \in {\bf Z} \, .  
\label{eq:ind-ell-appendix} 
\eea
\end{lem2}
\begin{proof}  
Applying relations (\ref{four-rec}) we show the following: 
\bea 
& & (n+1) ({  x}_{\ell}^{+})^{(n)} ({  x}_{1-\ell}^{-}(a))^{(n+1)} = 
 ({  x}_{\ell}^{+})^{(n)} \, {  x}_{1-\ell}^{-}(a) \, 
({  x}_{1-\ell}^{-}(a))^{(n)} \non \\
& = & {  x}_{1-\ell}^{-}(a) \, ({  x}_{\ell}^{+})^{(n)} 
({  x}_{1-\ell}^{-}(a))^{(n)} 
+ \underline{[({  x}_{\ell}^{+})^{(n)},  {  x}_{1-\ell}^{-}(a)]} \,  
({  x}_{1-\ell}^{-}(a))^{(n)}
\non \\
&= & {  x}_{1-\ell}^{-}(a) \, ({  x}_{\ell}^{+})^{(n)} 
({  x}_{1-\ell}^{-}(a))^{(n)} 
+
 ({  x}_{\ell}^{+})^{(n-1)} \, \underline{{  h}_1(a) \, 
({  x}_{1-\ell}^{-}(a))^{(n)}}
+  \underline{{  x}_{\ell+1}^{+}(a) \,  
({  x}_{\ell}^{+})^{(n-2)}}
 ({  x}_{1-\ell}^{-}(a))^{(n)} \non \\ 
&= & {  x}_{1-\ell}^{-}(a) \, ({  x}_{\ell}^{+})^{(n)} 
({  x}_{1-\ell}^{-}(a))^{(n)} 
+ \left\{ 
 ({  x}_{\ell}^{+})^{(n-1)} \, \underline{  
({  x}_{1-\ell}^{-}(a))^{(n)}\, {  h}_1(a)} 
+  ({  x}_{\ell}^{+})^{(n-1)} \,  
[{  h}_1(a), ({  x}_{1-\ell}^{-}(a))^{(n)}] \right\} \non \\
& & \qquad  + {\frac 1 2} [{  h}_1(a),  
({  x}_{\ell}^{+})^{(n-1)} ] \, 
 ({  x}_{1-\ell}^{-}(a))^{(n)}  \, . 
\label{eq:line}
\eea
Here, substituting the product
$({  x}_{1-\ell}^{-}(a))^{(n)} \, {  h}_1(a)$ by   
\bea  
& & [{  x}_{\ell}^{+}, ({  x}_{1-\ell}^{-}(a))^{(n+1)} ] 
+  {  x}_{2-\ell}^{-}((a)^2) \,  ({  x}_{1-\ell}^{-}(a))^{(n-1)}
\non \\
& = & {  x}_{\ell}^{+} \, ({  x}_{1-\ell}^{-}(a))^{(n+1)}  
+  {  x}_{2-\ell}^{-}((a)^2) \,  ({  x}_{1-\ell}^{-}(a))^{(n-1)}
- ({  x}_{1-\ell}^{-}(a))^{(n+1)} \, {  x}_{\ell}^{+} \, ,  
\non 
\eea
we show that the second term  
$({  x}_{\ell}^{+})^{(n-1)} 
({  x}_{1-\ell}^{-}(a))^{(n)}\, {  h}_1(a)$ 
in the last lines of (\ref{eq:line}) is equal to the following:   
$$ 
n \, ({  x}_{\ell}^{+})^{(n)} \, ({  x}_{1-\ell}^{-}(a))^{(n+1)} 
+ ({  x}_{\ell}^{+})^{(n-1)} \, {  x}_{2-\ell}^{-}((a)^2) \, 
 ({  x}_{1-\ell}^{-}(a))^{(n-1)} 
- 
({  x}_{\ell}^{+})^{(n-1)} \, 
 ({  x}_{1-\ell}^{-}(a))^{(n+1)} \, {  x}_{\ell}^{+} \, .  
$$
Thus, we have 
\bea 
& & \left\{ (n+1) - n \right\}  
({  x}_{\ell}^{+})^{(n)} \, ({  x}_{1-\ell}^{-}(a))^{(n+1)} \non \\ 
& = &  {  x}_{1-\ell}^{-}(a) \, ({  x}_{\ell}^{+})^{(n)} 
({  x}_{1-\ell}^{-}(a))^{(n)} 
+
 ({  x}_{\ell}^{+})^{(n-1)} \underline{ {  x}_{2-\ell}^{-}((a)^2) \, 
 ({  x}_{1-\ell}^{-}(a))^{(n-1)} }
\non \\
& & \qquad + 
({  x}_{\ell}^{+})^{(n-1)} \,  
[{  h}_1(a), ({  x}_{1-\ell}^{-}(a))^{(n)}] \non \\
& &  + {\frac 1 2} [{  h}_1(a),  
({  x}_{\ell}^{+})^{(n-1)} ] \, 
 ({  x}_{1-\ell}^{-}(a))^{(n)}
- 
({  x}_{\ell}^{+})^{(n-1)} 
 ({  x}_{1-\ell}^{-}(a))^{(n+1)} \, {  x}_{\ell}^{+} \, .  
\label{eq:line2}
\eea
Putting ${  x}_{2-\ell}^{-}((a)^2) \, 
({  x}_{1-\ell}^{-}(a))^{(n-1)} =
 - (1/2) [{  h}_1(a), ({  x}_{1-\ell}^{-}(a))^{(n)}] $ 
 into (\ref{eq:line2}) 
we have relation (\ref{eq:ind-ell-appendix}).   
\end{proof}

%
%

 \setcounter{equation}{0} 
 \renewcommand{\theequation}{C.\arabic{equation}}

\section{Reduction relations for $a \ne 0$}

Substituting $\ell=0$ in $({\rm B}_n)$ and making use of (\ref{eq:h-alpha}),  
we have the following: 
\begin{cor2}
\be 
\lambda_n(a) = {\frac 1 n} \sum_{k=1}^{n} (-1)^{k-1} d_k((a)^k) \, 
\lambda_{n-k}(a) \, 
 , \quad  \mbox{\rm for} \quad  n=1, 2, \ldots, r. 
\label{eq:dk-lambda-a}
\ee
\label{cor:dk-lambda-a}
\end{cor2}

\begin{lem2}
For any integer $\ell$ we have 
\be 
({  x}_{\ell}^{+})^{(k)} ({  x}_{1-\ell}^{-}(a))^{(k)} 
\Omega = \lambda_k(a)  \Omega \, , \quad  \mbox{\rm for} \quad  
k = 1, 2, \ldots, r . 
\label{eq:xx-lambda-a}
\ee
\label{lem:xx-lambda-a}
\end{lem2} 
\begin{proof} 
We show (\ref{eq:xx-lambda-a}) by induction on $k$. 
The $k=1$ case is shown directly. 
Assuming (\ref{eq:xx-lambda-a}) for $k \le n-1$,  
we derive the $k=n$ case 
by $({\rm B}_{n})$ of lemma \ref{lem:ABC},  
(\ref{eq:dk-lambda-a}) and (\ref{eq:h-alpha}).  
\end{proof}

\begin{prop2} 
We have  
\be 
{x}_{r+1-\ell}^{-}(a) \, \Omega = \sum_{j=1}^{r} (-1)^{r-j} 
\lambda_{r+1-j}(a) \, {x}^{-}_{j-\ell}((a)^j) \, \Omega \, ,
 \quad {\rm for} \, \, 
\ell \in {\bf Z} \, .  
\label{eq:rr-ell-a}
\ee
\label{lem:rr-ell-a}
Here we recall that 
$\lambda_j(a)$ is defined by (\ref{eq:lambda-a}) 
with the highest weight parameters. 
\end{prop2} 
\begin{proof}
We derive  reduction relation (\ref{eq:rr-ell-a}) from 
$({\rm A}_{r+1})$ of lemma \ref{lem:ABC} and lemma \ref{lem:xx-lambda-a}.  
\end{proof}

%
%

 \setcounter{equation}{0} 
 \renewcommand{\theequation}{D.\arabic{equation}}
\section{Proof of eqs. (\ref{eq:sum}) of lemma \ref{lem:submodule} } 

In order to derive the first relation of (\ref{eq:sum}),  
we first show the following: 
\be
{[} h_n , w_{j^{k}}({\hat {\bm a}}) {]} = 
 (-2) \sum_{t=0}^{n} 
\left( 
\begin{array}{c} 
n \\
t
\end{array} \right) 
 \, a_j^{n-t} 
w_{j^{k-t}}({\bm {\hat {a}}})  \quad \mbox{for} \, \, 
n > 0.  \label{eq:sumA} 
\ee
We show it by induction on $n$. For $n=1$ we have 
\bea {[} h_1, w_{j^k}({\bm {\hat a}}) {]}/(-2) 
& = & x_{r-k+1}^{-}({\bm  {\hat a}} \setminus j^{k})  \non \\  
& = &  x_{r-k+1}^{-}({\bm {\hat a}} \setminus j^{k-1}) + 
a_j \, x_{r-k}^{-}({\bm {\hat a}} \setminus j^{k})  \non \\ 
& = & w_{j^{k-1}}({\bm {\hat a}}) + a_j w_{j^k}({\bm {\hat a}}) 
  \, . \non  
\eea
Let us assume (\ref{eq:sumA}) in the case of $n$. 
In order to derive (\ref{eq:sumA}) for the case of $n+1$, 
we first make use of the following: 
\bea 
{[} h_{n+1}, w_{j^k}({\bm {\hat a}})  {]} /(-2) 
& = & x_{r-k+n+1}^{-}({\bm {\hat a}} \setminus j^{k}) \non \\ 
 & = & x_{r-(k-1)+n}^{-}({\bm {\hat a}} \setminus j^{k-1}) 
  + a_j \, x_{r-k+n}^{-}({\bm {\hat a}} \setminus j^{k}) \, , \non \\ 
& = & {[} h_n, w_{j^{k-1}}({\bm {\hat a}}) {]} 
  + a_j \, {[} h_n, w_{j^{k}}({\bm {\hat a}}) {]} \, . \label{eq:h(n+1)} 
\eea
Substituting relation (\ref{eq:sumA}) for $n$ 
into (\ref{eq:h(n+1)}) and making use of the recursive relation: 
 $_{n+1}C_t = _nC_{t} + _nC_{t-1}$, 
we obtain relation (\ref{eq:sumA}) for the case of $n+1$.  

We now discuss 
the second relation of (\ref{eq:sum}).   
We show it by induction on $n$ and $k$. 
We first derive it for the case of $n=1$ and for arbitrary $k$ with 
$k \ge 1$. Through induction on $k$, it is easy to show the following: 
\be 
{[} h_{-1}, w_{j^{k}}({\hat {\bm a}}) {]} \, \Omega =  
(-2) \sum_{t=0; t < k} 
\left( 
\begin{array}{c} 
-1 \\
t
\end{array} \right) 
 \, a_j^{-1-t} 
w_{j^{k-t}}({\hat {\bm a}}) \Omega \, . 
\ee 
Assuming the case of $n$ and $k$,  
we now show the case of $n+1$ and $k$. 
We first note 
\be 
x_{r-m-n-1}^{-}({\bm {\hat a}}/j^{k}) = 
a_j^{-1} x_{r-m-n}^{-}({\bm {\hat a}}/j^{k}) 
- a_j^{-1} x_{r-m-n}^{-}({\bm {\hat a}}/j^{k-1}) 
\ee
and make use of the following relation: 
\be 
\left( 
\begin{array}{c} 
-n-1 \\
k
\end{array} 
\right)
= \left( 
\begin{array}{c} 
-n \\
t 
\end{array} 
\right)
- \sum_{\ell=1}^{k} 
\left( 
\begin{array}{c} 
-1 \\
\ell -1 
\end{array} 
\right)
\left( 
\begin{array}{c} 
-n \\
k-\ell 
\end{array} 
\right)  \, , 
\ee
we obtain the second relation of (\ref{eq:sum}) for 
the case of $n+1$ and $k$.

\end{document}